\documentclass[11pt,preprint]{aastex}

\shorttitle{RR Lyraes in M3}
\shortauthors{Cacciari, Corwin \& Carney}

\begin{document}


\title{A multi-color and Fourier study of RR Lyrae variables in the globular 
cluster NGC 5272 (M3)}


\author{C. Cacciari\altaffilmark{}}  
\affil{Osservatorio Astronomico, Via Ranzani 1, 40127 Bologna, Italy}
\email{cacciari@bo.astro.it}

\author{T.M. Corwin\altaffilmark{1}}  
\affil{Department of Physics, University of North Carolina,
    Charlotte, NC 28223}
\email{mcorwin@uncc.edu}

\and

\author{B.W. Carney\altaffilmark{1}}
\affil{Department of Physics and Astronomy, University of North Carolina, 
   Chapel Hill, NC 27599}
\email{bruce@astro.unc.edu}


\altaffiltext{1}{Visiting Astronomer, Kitt Peak National Observatory, 
National Optical Astronomy Observatories, which are operated by AURA, Inc.\ 
under cooperative agreement with the National Science Foundation.}


\begin{abstract}
We have performed a detailed study of the pulsational and evolutionary
characteristics of 133 RR Lyrae stars in the globular cluster
NGC5272 (M3) using highly accurate BVI data taken on 5 separate epochs.  
M3 seems to contain no less than $\sim$32\% of Blazhko stars, and the 
occurrence and characteristics of the Blazhko effect have been analyzed in 
detail. 
We have identified a good number ($\sim$ 14\%) of overluminous RR Lyrae stars 
that are likely in a more advanced evolutionary stage off the Zero Age 
Horizontal Branch (ZAHB). \\
Physical parameters (i.e. temperature, luminosity, mass) have been derived 
from (B--V) colors and accurate color-temperature calibration, and compared 
with Horizontal Branch evolutionary models and with the requirements of 
stellar pulsation theory.   
Additional analysis by means of Fourier decomposition of the V light curves 
confirms, as expected, that no metallicity spread is present in M3. Evolution 
off the ZAHB does not affect [Fe/H] determinations, whereas Blazhko stars 
at low amplitude phase do affect [Fe/H] distributions as they appear 
more metal-rich. Absolute magnitudes derived 
from Fourier coefficients might provide useful average estimates for groups of 
stars, if applicable, but do not give reliable {\em individual} values. 
Intrinsic colors derived from Fourier coefficients show significant 
discrepancies with the observed ones, hence the resulting temperatures 
and temperature-related parameters are unreliable.  

\end{abstract}




\section{Introduction}

M3 (NGC5272: RA[2000] = 13:42:11, DEC[2000] = 
+28:22:32) is one of the most important globular clusters of the 
Galactic halo. It is located at $\sim$ 11.9 kpc from the Galactic 
centre, 10.0 kpc from the Sun and 9.7 kpc above the Galactic plane 
(Harris 1996). 
It contains by far the largest number of variable stars (Bakos et al.\ 
2000 assign numbers to 274 variables) with a rather high 
specific frequency (i.e. normalized to total mass) of RR Lyrae stars. 
Its metallicity, for a long time considered to be [Fe/H] = --1.66 
(Zinn \& West 1984, hereafter ZW), has been revised using high resolution 
spectroscopic data of giant stars and appears to be somewhat 
higher, i.e. --1.47 (Kraft et al.\ 1992) or --1.50 (Kraft \& Ivans 2003, 
hereafter KI) in the Fe$_{II}$ metallicity scale, and  --1.34 
(Carretta \& Gratton 1997 in their own metallicity scale).  

Since the seminal paper by Oosterhoff (1939), that subdivided the globular
clusters in two groups according to the mean period of their RRab 
variables ($<P_{ab}>$=0.55d and 0.65d, respectively), 
M3 has traditionally 
been considered the  prototype Oosterhoff type I (OoI) cluster. 
The pulsational and evolutionary properties 
of its RR Lyrae variables were used by Sandage et al.\ (1981) 
as a reference template for the 
OoI group to compare to the RR Lyrae properties in other 
clusters, in particular those belonging to the Oosterhoff II 
(OoII) group. On this basis Sandage et al.\ (1981) reached the conclusion 
that there was a systematic period-shift (at fixed light curve 
amplitude) as a function of metallicity with respect to 
the assumed template relation (i.e. M3) in the $\log\,P - A_B$ plane; 
this effect was interpreted as due to a difference in luminosity, 
in the sense that longer period (more metal-poor) variables were 
intrinsically brighter. Although the subsequent interpretations  
of the period-shift effect were rather controversial and 
involved the r\^ole played by several other factors (Lee et al.\ 1990; 
Carney et al.\ 1992, hereafter CSJ; Sandage 1993, Clement \& Shelton 1999a,b),  
and the very existence of the period-shift was sometime questioned 
(e.g. Brocato et al.\ 1996), the position of M3 as {\em the} template OoI 
cluster was never questioned.  

However, the data for the RR Lyrae variables used in all these analyses 
were still essentially the UBV photographic 
observations by Roberts and Sandage (1955), Baker and Baker 
(1956) and Sandage (1959). 
A large and detailed morphological study of the light curve 
characteristics for 113 variables was performed by Szeidl (1965, 
1973), using photographic material taken over a long time baseline. 
It is only in the late '90s that systematic studies of the 
RR Lyrae variables in M3 with CCD detectors were undertaken, 
leading to large very accurate BVI photometric databases: 
Kaluzny et al.\ (1998, hereafter Kal98) with V data for 42 stars; 
Carretta et al.\ (1998, hereafter Car98) with BVI data for 60 stars; 
and Corwin \& Carney (2001, hereafter CC01) with BV data for 207 stars. 
These data offer the unprecedented opportunity to study the largest number 
of RR Lyrae stars ever detected within a single globular cluster, using 
well defined light curves obtained in more than one color and at different 
epochs. 

In this paper we present a detailed study of the pulsational 
and evolutionary characteristics of 133 RR Lyrae variables in M3, 
selected among those with the best quality light curves. 
We describe in Sect. 2 the data we have used, and we introduce  
the Blazhko effect and the definition of mean magnitudes and colors. 
In Sect. 3 we present the main relations between period, amplitude, magnitude 
and color, and in Sect. 4 we derive the reddening and discuss the calibration 
of temperatures and magnitudes. In Sect. 5 we derive the physical parameters of
the stars and compare them with evolutionary models, and discuss specifically 
the stars that are in a more advanced stage of evolution. Finally, 
in Sect. 6 we present the Fourier analysis of the light curves and discuss 
the implications on the estimate of the stellar physical parameters.  
A summary and the conclusions of this analysis are given in Sect. 7.

\section{The data}

The data sets we have used for the present analysis are:

\noindent $\bullet$ 
The BV data described in detail by CC01. 
These consist of 83 pairs of BV frames taken in May 1992,  102 pairs 
taken in April 1993, and 5 V and 4 B frames taken in June 1997, on a total 
of 207 variable stars.  This is the main data set we have used for our 
analysis in the B and V bandpasses.  

\noindent $\bullet$
The BVI data described by Car98. These 
consist of 65 to 69 frames in each color, taken in March 1990 and 
in Feb-Apr-May 1992, on a total of 60 RR Lyrae variables. 
This data set is the only one that provides information on the I 
bandpass. The B and V data have been used mainly to support and 
provide complementary information to the CC01 data.    

\noindent $\bullet$
The data described by Kal98. These consist of 
176 V frames taken in March-April 1996, on a total of 42 RR Lyrae 
variable stars. Also these data were used to support and 
provide complementary information to the CC01 data.

\subsection{Comparison with previous data sets}

A detailed comparison of the intensity-averaged magnitudes 
from the CC01 data set with  Car98, using common 
non-Blazhko variable stars, shows that CC01 V magnitudes are brighter 
than Car98 V photometry in the EAST and WEST fields by 0.065 
and 0.020 mags, respectively. As for the B magnitudes, CC01 photometry is 
brighter than Car98  EAST and WEST fields by 0.107 and 0.025 mags, 
respectively. 
We remind the reader that Car98 EAST field 
contains the variables n. 10, 31, 32, 34, 43, 57, 68, 69, 70, 
74, 75, 78, 84, 87, 100, 101, 128, 146, 149, 150, 178 and 197, 
and the WEST field contains the variables n. 6, 24, 25, 27, 28, 29, 
30, 41, 42, 46, 47, 58, 66, 67, 76, 77, 88, 109, 110, 111, 121, 
129, 130, 131, 132, 133, 134, 135, 136, 140, 142, 143, 155, 167, 
168, 170, 188 and 209. 
Finally, Car98 provided also I photometry with the warning that it may be 
affected by a zero-point error in the absolute calibration. We treat this
problem in some detail in Sect. 4.1, but we anticipate here that 
indeed Car98 I data are most likely too faint by $\sim$ 0.083 mag. \\ 
The same type of comparison with Kal98 data, using 
intensity integrated $<V>$ magnitudes for both data sets (note however 
that the published Kal98 $<V>$ are magnitude integrated),  
shows that the CC01 V magnitudes are brighter than Kal98 V photometry 
in the SOUTH and NORTH fields by 0.028 and 0.018 mags, respectively. \\
Car98 and Kal98 data have been corrected by the above 
offsets, when they have been used along with CC01 data (e.g. 
for Blazhko stars). \\
On the other hand, a comparison of CC01 photometry with 15 
randomly selected secondary standard stars from Sandage (1970) 
shows that CC01 V magnitudes are fainter by 0.009 $\pm$ 0.024 mag, 
and the B magnitudes are brighter by 0.004 $\pm$ 0.014 mag, as already 
noted by CC01. 

Only the stars with well defined light curves in both B and V  
bands have been taken into account for the present study. 
This led us to consider a total of 133 stars out of the 201 RR Lyraes
observed by CC01, in particular 23 RRc out of 43, 67 RRab out 
of 111, and 43 Blazhko stars out of 47. The stars we have not 
considered in the present study all have very noisy light 
curves, which may be due to photometric errors (contamination
from companions) or to intrinsic phenomena such as double-mode 
pulsation or unidentified Blazhko modulation. 
They may be very interesting objects in themselves, 
and surely deserve further and more careful investigation (cf.  
Clementini et al.\ 2004). 
However, for the purpose of the present analysis, we shall 
use only those stars that show the ``cleanest'' light curves 
so as to keep the noise at the minimum level, taking advantage 
of the fact that M3 is probably the only cluster where one 
can afford to be very selective, due to the richness of its 
variable star population.

\subsection{The Blazhko variables} 

The Blazhko effect, first noticed and studied by Blazhko (1907),  
is a modulation of the basic pulsation variability that produces  
variations of the light curve shape showing as larger photometric scatter 
and significant changes in the light curve amplitude. The timescale  
of this modulation is typically  tens of days but can be as large as 
a few hundred days. Several mechanisms have been proposed to explain the  
origin of this phenomenon, which however is still an open question.
We refer the reader to Smith (1995) for a recent and comprehensive 
discussion on this topic. 

According to CC01 data no less than $\sim$32\% of the total RR Lyrae variable 
star population in M3 is affected by Blazhko variability.  
This fraction might easily be larger if some of the stars with noisy light 
curves, that we have not considered in the present analysis, turn out to be 
Blazhko variables in future studies. The frequency of this phenomenon we 
find in M3 is consistent with previous results in other stellar systems 
(cf. Smith 1995) and has been recently confirmed in another cluster, 
NGC3201, where Piersimoni et al.\ (2002) have identified about 30\% such stars.  
However, the detection of Blazhko variability is not straightforward and 
needs monitoring the variables over several epochs, therefore it can easily 
go undetected even in relatively well studied clusters. 

The presence of Blazhko stars within a population of regular   
RR Lyrae variables may affect the global characteristics of this 
population, for example in all cases where the light curve amplitude 
is involved (e.g. the period-amplitude or color-amplitude relations). 
It certainly produces a scatter in the relations among various parameters, 
possibly  masking other subtle effects by drawning them into the noise. 
Therefore, {\em the knowledge of the Blazhko star population is essential 
in order to select a pure sample of regular stars that define the average 
characteristics of the cluster variable star population}. 

In the following sections we shall investigate these effects and try to 
derive the main characteristics of the regular variable population.

\subsection{Mean Magnitudes and Colors}

Mean magnitudes $<B>$ and $<V>$ of the variables have been derived by 
integrating the light curves in intensity and converting the result of this 
integration to magnitudes. We have not tried to correct these values to the 
equivalent static values, as proposed by Marconi et al.\ (2003), since 
the amplitude-dependent correction to apply e.g. to $<V>$ would be at most 
--0.02 mag at A$_V$=1.4 and all our stars have smaller amplitudes.  
We list the values of $<B>$ and $<V>$ 
thus derived for the RRc, RRab and Blazhko stars in Tables \ref{cfot}, 
\ref{abfot} and \ref{blafot}, 
respectively, as well as the B and V light curve amplitudes $A_B$ and 
$A_V$, and the corresponding rise time values ($RT_B$ and $RT_V$) defined 
as the phase intervals between the minimum and maximum B and V light, 
respectively. 
For the RRab and Blazhko stars we have calculated also the 
magnitude-integrated $B_{min}$ and $V_{min}$ at minimum light, 
i.e. $0.5 \le \phi \le 0.8$. 
These parameters have been derived with the help of Fourier 
decomposition of the light curves, using 6 harmonics for the 
RRc-type and 6 to 15 harmonics for the RRab-type variables.
For the Blazhko stars we have considered separately the CC01 data taken 
in 1992 and 1993, that provide different epochs for the study of the Blazhko 
phenomenon.  

The mean colors of the RR Lyare stars are a more controversial issue.
The choice of which mean color best reproduces the color the star would 
have were it not pulsating is a long standing problem, and has been discussed 
by several authors (cf. Silbermann \& Smith 1995 and references therein for 
a detailed review and discussion of this topic). 
Very briefly, several solutions have been proposed as ``best mean color'', 
e.g. the magnitude-averaged (B--V) (Preston 1961; Sandage 1990), or the 
intensity-averaged $<B>-<V>$ (Davis \& Cox 1980), or a combination such as 
$2/3<B-V>+1/3(<B>-<V>)$ (Lub 1977), or $<B>-<V>$+C(A) where C(A) is an 
empirical correction for amplitude (Sandage 1990). CSJ  
argued that, no matter how the average is done, the (B--V) colors are 
poor temperature indicators because they are distorted by surface gravity 
and non-LTE effects during a non negligible fraction of the pulsation cycle 
around maximum light, that can produce excess emission 
in the B band. They proposed a few formulae to estimate the temperature,
involving the  (B--V) colors plus corrective factors due 
to metallicity or amplitude or period, and also a formula involving only 
period, amplitude and metallicity (cf. their Eq.s 13-16). 
On the theoretical side, Bono et al.\ (1995) have calculated synthetic mean
colors for convective pulsating models over a wide range of luminosities and 
temperatures, and found that indeed both (B--V) and  $<B>-<V>$ colors differ 
from the equivalent static color by a quantity that is a function of the 
light curve amplitude. The corrections they derive (their Table 4)         
are generally similar to the empirical corrections estimated by Sandage 
(1990). 

Given the nearly unanimous consensus that $<B>_{int}-<V>_{int}$ colors plus 
some sort of amplitude-related correction reproduce reasonably well the 
equivalent static colors - that we shall call $(B-V)_S$, we adopt this solution 
where the corrections are those estimated by Bono et al.\ (1995).  
The values of $(B-V)_S$ are listed in Tables \ref{cfot} and \ref{abfot}, 
where  we report for convenience also the (B--V)$_{mag}$ colors of the RRc and 
RRab stars taken from CC01, that were calculated as $<B>_{mag}-<V>_{mag}$.   
For a consistency check, we have calculated the values of   
$(B-V)_S$ both from (B--V) and $<B>-<V>$ colors and applying the corresponding 
amplitude corrections, and we have verified that the results agree 
within 0.02 mag, and mostly within 0.01 mag. Also the use of 
Marconi et al.\ (2003) formulation (cf. their eq. 16) 
to derive the static (B--V) colors produces similar values to $(B-V)_S$ 
within 0.01 mag.  We estimate that typical internal errors of the average 
$<B>$ and $<V>$ magnitudes and $(B-V)_S$ colors are 0.01 and 0.02 mag, 
respectively.  
  
For the Blazhko stars listed in Table \ref{blafot} we have reported no mean 
colors from CC01, because they were obtained from the combined 1992 and 1993 
data sets  and  have lost information on the Blazhko phase. Instead, 
the $(B-V)_S$  colors for the two epochs separately can be 
derived from the corresponding $<B>_{int}$ and $<V>_{int}$ average magnitudes, 
and are listed  in Tab. \ref{blafot}.


\section{The main relations between Period, Amplitude, Rise Time, Magnitude 
and Color} 

These basic parameters of the RR Lyrae variable stars are 
connected by relations that reveal important physical, evolutionary and 
pulsational characteristics. Most of the considerations that 
we present below were already outlined by CC01, but we repeat them here in 
more detail and for the sake of convenience in the following analysis. 

\subsection{The Color-Magnitude diagram}

\begin{figure}
\plotone{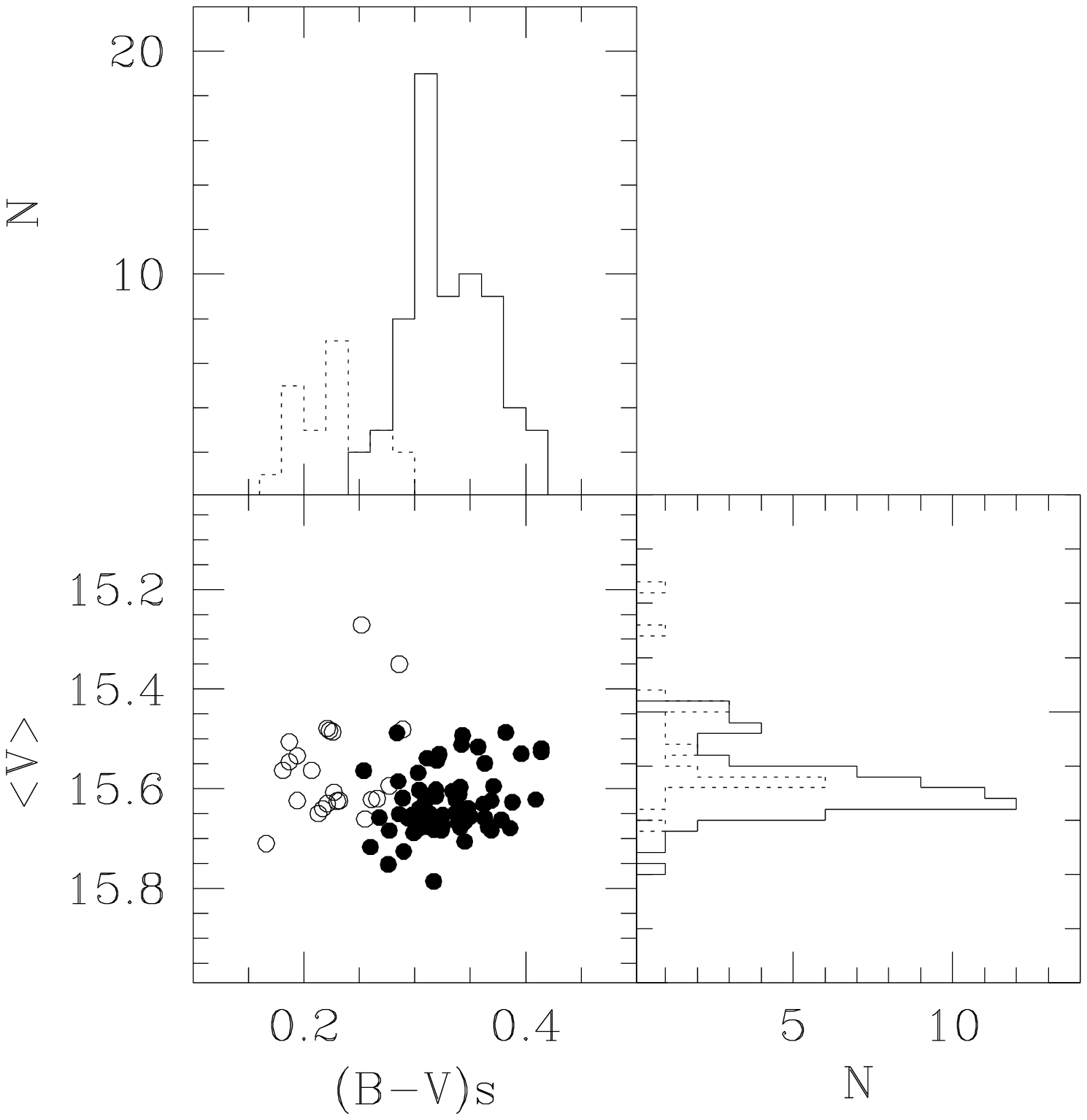}
\caption{Color-Magnitude diagram for the  
c-type (open circles) and ab-type (filled circles)  RR Lyrae 
variables listed in Table 1 and 2, respectively (lower left panel). 
The histograms of the color distributions (upper panel) and the $<V>$ 
magnitude distributions (lower right panel) for the RRc (dotted line) and the 
RRab (solid line) are also shown. 
\label{cacciari.fig1}}
\end{figure}

The Color-Magnitude diagram of the full sample of 207 variable 
stars  has been discussed already by CC01. 
We present in Fig. \ref{cacciari.fig1} a less populous but cleaner version based 
on the sub-sample of RRc and RRab stars studied in this paper, using the 
$<V>$ and $(B-V)_S$ values listed in Tables \ref{cfot} and \ref{abfot}. 
The main characteristics already discussed by CC01 are here reconfirmed, 
namely: \\
i) the blue and red edges of the color distribution are located at 
$(B-V)_S$=0.18 and 0.42 mag, respectively. Note, however, that one star (V178) 
is slightly bluer than this limit [$(B-V)_S$=0.166], and is the faintest 
of the entire group of RRc variables. \\
ii) There is overlap in color between RRc and RRab variables, that occurs in 
the interval $\sim$0.24 to 0.30. As CC01 noted, it is possible to draw a 
line slanting toward redder colors at brighter magnitudes that separates 
most of the RRab from the RRc variables, but still two unusually bright 
RRab stars (V42 and V96) fall in the RRc area. 
Whereas V96 might be dismissed because it has a somewhat incomplete light 
curve, V42 does not show any special problem in the photometry, and  its 
unusual position in the CMD seems to be due to the combination of unusually 
bright $<V>$ and large amplitude color-correction. \\
iii) The Blazhko stars (not shown to avoid confusion) overlap the area of the 
RRab stars avoiding however the reddest part of the color distribution: they 
only reach as red as $(B-V)_S \le$ 0.39 with CC01 data. \\
iv) The magnitude distribution (lower right panel of Fig. \ref{cacciari.fig1}) shows 
that the main body of the RRab population peaks around 
$<V>$=15.64$\pm$0.04 mag, and there is a clear separate group of 12 stars 
(V26, 31, 42, 48, 58, 60, 65, 104, 124, 146, 186, and 202) at brighter 
magnitudes (all individual $<V>\le$15.56, centered at $<V>\sim$15.52$\pm$0.02). 
Most of these stars were already noted as very luminous by CC01.  
The main distribution of the RRc stars seems to be shifted by $\sim$0.05 mag 
towards brighter magnitudes ($<V>$=15.59$\pm$0.06 mag), and also shows 
a tail of very luminous stars, in particular six (V29, 70, 85, 129, 140, 
and 170) that are all brighter than 15.5. 
We do not see any statistically significant evidence of {\em four} populations 
from the $<V>$ distribution, as claimed by Jurcsik et al.\ (2003). \\
v) On the faint end, the distribution appears to end statistically at 
$<V>\sim$15.72, as was found also by CC01; this value is taken to 
correspond to the lower envelope of the Horizontal Branch (HB) luminosity 
distribution, i.e. the Zero Age Horizontal Branch (ZAHB). 
Only 3 RRab stars are fainter than this value. 
Based on the $<V>$ histogram of the RRab stars we note that the thickness of 
the HB is $\le$0.20 mag or $\sim$0.30 mag, depending on whether we exclude or 
include the brighter stellar component (cf. Sandage \& Katem 1982).  \\
vi) The $<V>$ distribution of the Blazhko stars (not shown) reaches about 
the same edges of the distribution of the normal stars, but is skewed towards 
the brighter magnitudes. A group of seven stars (V3, 14, 24, 44, 78, 130 and 
143) are as bright as the more luminous group of regular stars identified 
in item iv), and in fact their average magnitude is $<V>$=15.52$\pm$0.04. 
The main population of Blazhko stars has $<V>$=15.65$\pm$0.05 which is 
basically identical to the mean value we find for the regular RRab variables. 
We note that the average magnitude of the Blazhko stars in the small-amplitude 
Blazhko phase appears to be $\sim$0.02 mag fainter than in the 
large-amplitude phase, but this difference is hardly significant from the 
statistical point of view. 
      
\subsection{The Period-Amplitude and Period-Rise Time diagrams} 

\begin{figure}
\plotone{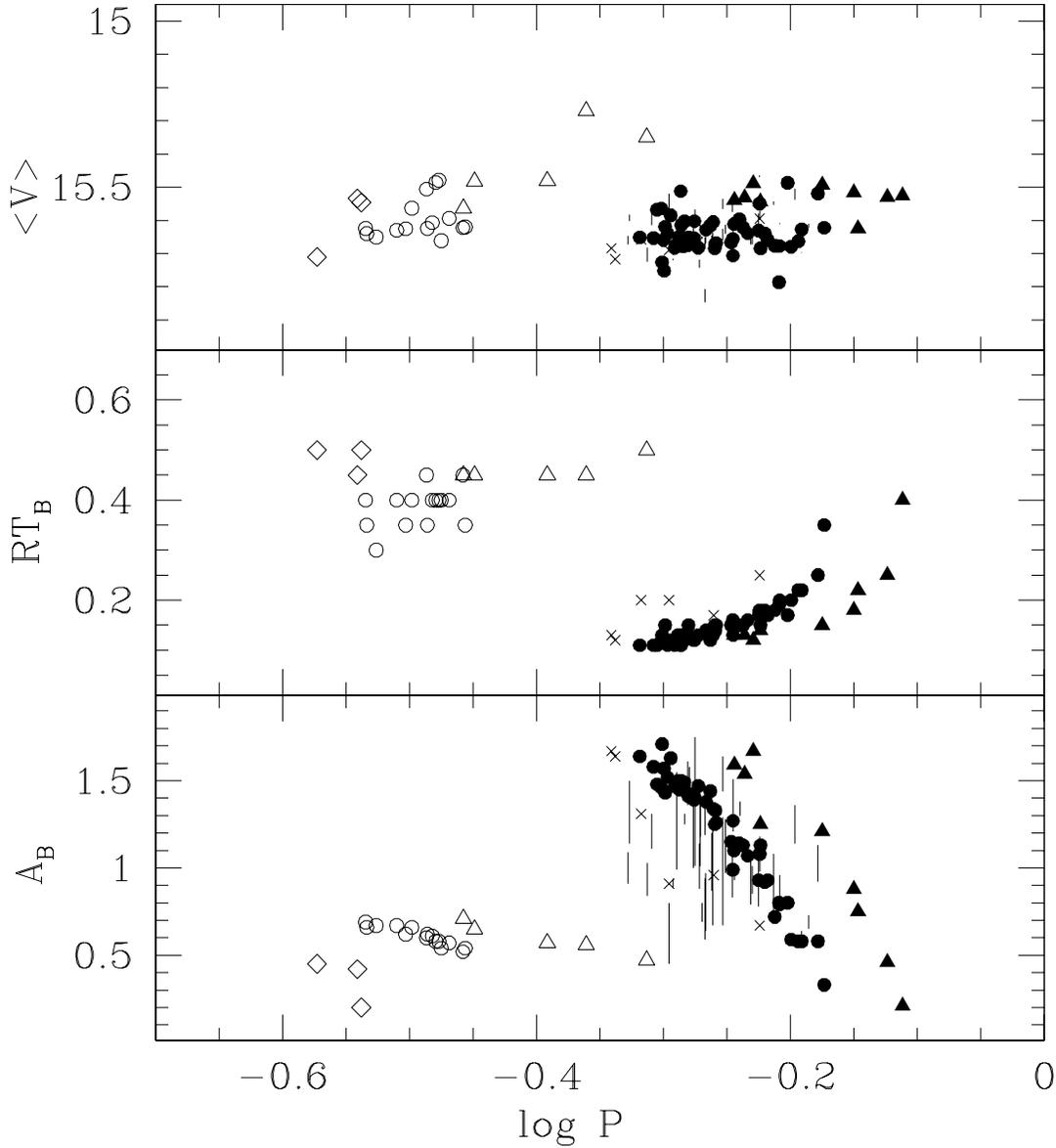}
\caption{Lower panel: blue light curve amplitude $A_B$ vs. $\log\,P$ 
for the c-type (open symbols) and ab-type (filled symbols) RR Lyrae 
variables listed in Table \ref{cfot} and \ref{abfot}, respectively. 
The stars that are likely evolved off the ZAHB, and show up at longer 
period for a fixed amplitude, are indicated as triangles. The crosses 
indicate the suspected Blazhko stars, and the known Blazhko stars are 
shown as lines connecting the 1992 and 1993 data sets, from Table \ref{blafot}. 
The short-period small-amplitude RRc variables V105, V178 and V203 are shown 
as diamonds. 
Middle panel: same as lower panel, for the rise time values of the B 
light curves. No Blazhko stars are shown, because the rise times are quite 
uncertain for these stars.  
Upper panel: the corresponding $<V>$ values. 
\label{cacciari.fig2}}
\end{figure}

As has long been known, the amplitudes of the RRab stars 
are strongly correlated with period, whereas the amplitudes of the RRc stars 
have a  nearly flat distribution. A similar behaviour is shown by the 
rise time. 
We show in Fig. \ref{cacciari.fig2}  the amplitudes A$_B$ (lower panel) and the rise 
times $RT_B$ (middle panel) of the blue light 
curves vs. period ($\log\,P$), and for ease of discussion the 
corresponding values of $<V>$ (upper panel). 

\subsubsection{The RRc variables}
 
The main body of the RRc distribution defines a clear nearly flat 
sequence in the A$_B$ vs. $\log\,P$  plane. 
Then there is a group of 3 stars at 
the short period end of this distribution, with particularly small amplitudes, 
and another group of 5 stars that seem to define a nearly parallel 
distribution to the main one, but shifted to larger amplitudes and/or 
longer periods. These stars also stand out in the $RT_B$ vs. $\log\,P$  plane.
In more detail: \\
i) The three short-period small-amplitude RRc stars are V105, V178 and V203
(shown as diamonds in Fig. \ref{cacciari.fig2}). 
They have normal values of $<V>$, including V178 that has the shortest 
period and the faintest magnitude  but its $<V>$ is still compatible with 
the general trend of $<V>$ vs period among the RRc stars. They seem to be 
a normal extension of the main RRc population, according to theoretical models 
(Bono et al.\ 1997) and observational evidence also in other clusters 
(Clement \& Shelton 1999b), showing that first overtone pulsators do have  
decreasing amplitudes at the short period (high temperature) end of the 
distribution, i.e. a bell-shaped distribution.     
On the other hand, stars with periods shorter than $\sim$0.29 day 
($\log\,P\sim$--0.54) and sinusoidal light curves with particularly small 
amplitudes have been found 
to exist also in several other globular clusters, as well as in the 
large sample of field LMC variables from the MACHO survey (Alcock et al.\ 1996) 
and in the Galactic Bulge (Olech 1997), where they show a well-defined peak 
in the period distribution and thus may define a separate population of 
variable stars.  These stars could be second overtone (RRe) pulsators  
(see Clement \& Rowe 2000 for references and a detailed discussion of this 
topic).  
Thus, our three short-period small-amplitude RRc stars could be either second 
overtone or regular first overtone pulsators.  
A way to test the pulsation mode of V105, V178 and V203 is to plot 
the Fourier parameters $\phi_{21}$ vs. $A_{21}$ of their light curves, 
as we have done in Sect. 6.1. Anticipating the results presented there, 
we suggest that only V203 is a likely RRe star, 
whereas V105 and V178 seem to be regular RRc stars. \\  
ii) The group of five RRc stars defining the sequence at  larger 
amplitudes and/or longer periods includes V70, V85, V129, V170 and V177 
(shown as open triangles).  They were already noticed in the previous section 
item iv) for being significantly brighter than the main body of the 
RRc distribution ($<V>$=15.43$\pm$0.12).  We consider 
these stars as belonging to a category that we shall call  
``longP/overluminous''  stars.  

\subsubsection{The RRab variables} 

\begin{figure}
\plotone{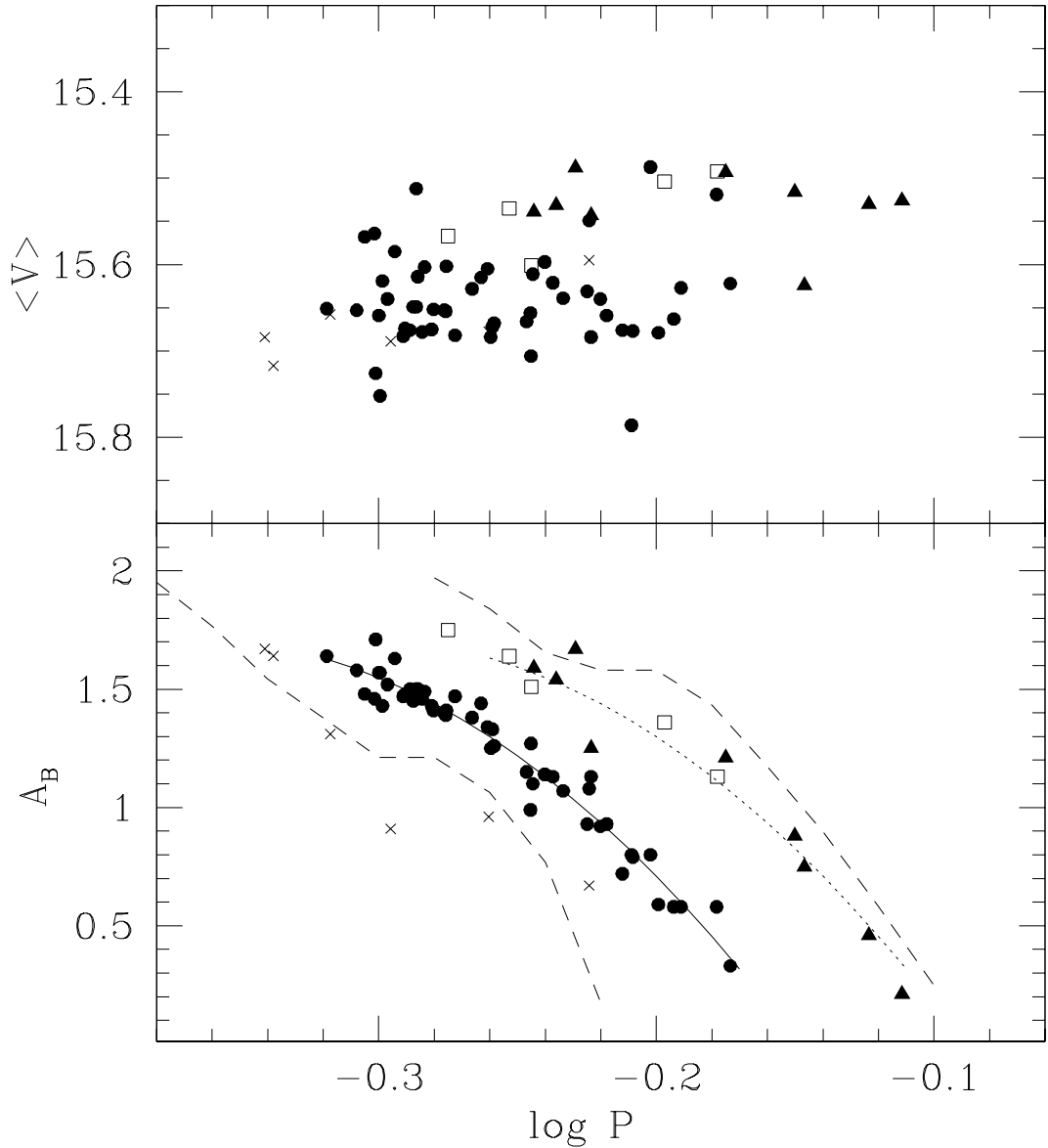}
\caption{A blow-up of Fig. 1 for the RRab (and some Blazhko) stars only. 
Upper panel: same as upper panel in Fig. \ref{cacciari.fig2}. 
Lower panel: the solid curve represents the mean distribution of the 
{\it bona fide} regular stars (filled circles).  The dotted curve is obtained 
by shifting the solid curve to longer periods by $\Delta \log\,P$=+0.06, and 
represents quite well the long-period, candidate evolved RRab stars (filled 
triangles). We show for completeness also the candidate evolved Blazhko stars 
at the phase of maximum amplitude (open squares). Stars with smaller 
than normal amplitudes (suspected Blazhko) are shown as crosses. 
Note that the best representation of these distributions is not a linear 
but a quadratic relation. 
We show for comparison the theoretical relations calculated by Piersimoni 
et al.\ (2002) for $\log\,L$=1.61 (left dashed line) and 1.72 
(right dashed line).  
\label{cacciari.fig3}}
\end{figure}

Among the RRab stars, one sees that the large scatter in amplitude 
is closely mirrored by a scatter in $<V>$ and appears also in the $RT_B$ 
parameter. We show in Fig. \ref{cacciari.fig3} a 
blown up version of Fig. \ref{cacciari.fig2} for the sake of clarity, and we see   
that the group of the RRab variables also appears to be made of three 
subgroups:  \\
i) The main body of the RRab distribution,  defining the well known 
relation between $\log\,P$ and $A_B$. This relation  has traditionally 
been taken as  a linear approximation; however, we see that the distribution 
for our stars is better represented by a quadratic relation 
$A_B = -3.123-26.331\log\,P-35.853\log\,P^2$, r.m.s. error of the fit 
$\sigma\sim$ 0.08. Also in the $\log\,P-RT_B$ plane a quadratic relation 
(e.g. $RT_B = 0.781 +4.269\log\,P + 6.881\log\,P^2$, r.m.s. error of the 
fit $\sigma \sim$ 0.02) seems to provide a better fit of the data. 
Theoretical models calculated for $\log\,L$=1.61 and 1.72 and values of 
mass and metallicity quite adequate for M3 (Piersimoni et al.\ 2002; 
Marconi et al.\ 2003) are also  reported in Fig. \ref{cacciari.fig3} 
(lower panel): they have a similar non-linear shape to our distributions 
(a part from a  ``hump'' in the middle range that is not quite so evident in 
the data).   \\
ii) A group of 6 stars (V22, 54, 71, 72, 77 and 144, shown as crosses) 
at shorter periods or, more likely, smaller amplitudes (and larger than 
normal $RT_B$ values, i.e. a different shape of the light curve) than the 
main body of RRab stars, with  $<V> = 15.67 \pm 0.04$. These stars are 
compatible with being {\em unrecognized} Blazhko variables observed during 
the low-amplitude phase of the Blazhko modulation, as we discuss below. 
We shall call them for convenience ``low amplitude/suspected Blazhko'' 
stars.  \\
iii) A group of 9 stars (V26, 31, 42, 60, 65, 104, 124, 202 and KG14, shown as 
filled triangles) with longer periods at a given amplitude. 
All of them except one (KG14) were already noticed in the previous section 
item iv) for being significantly brighter than the main body of the RRab 
distribution. They have $<V> = 15.53 \pm 0.04$ mag 
and can be considered to belong to the longP/overluminous group.
Some of these stars (those with the longest period) stand out from the main 
distribution also in the $RT_B$ vs. $\log\,P$ plane.  
These stars seem to be well represented by the same quadratic relation defined 
by the RRab stars, shifted toward longer periods (at fixed amplitude) by 
$\Delta\log\,P \sim$ 0.06. This shift corresponds approximately to the mean 
location traditionally assigned to OoII variables (cf. Sandage et al.\ 1981). 
Although a detailed discussion of the Oosterhoff dichotomy is beyond the 
scope of this paper, we exploit our beautiful sample to investigate briefly 
a few basic issues related to the period-shift effect in the next section. 

\subsubsection{The period-shift effect}

\begin{figure}
\plotone{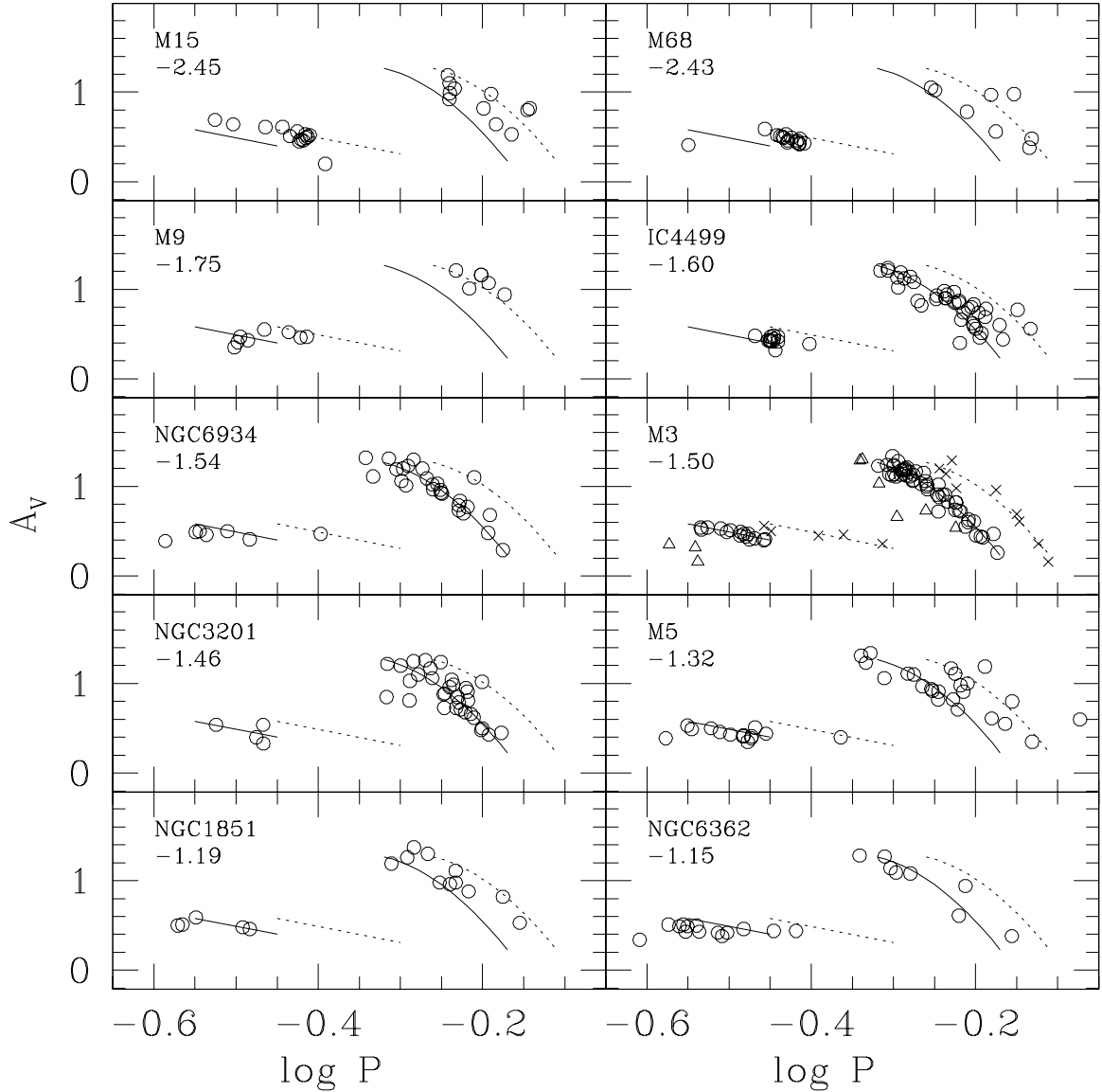}
\caption{Period-amplitude distributions for RRc and RRab variables in three 
OoII clusters (M15, M68 and M9), three intermediate type clusters (IC4499,
NGC6934 and NGC1851), and three OoI clusters (NGC3201, M5 and NGC6362), 
compared to M3. The mean distributions of the M3 regular (solid lines) 
and evolved (dotted lines) stars are shown in each panel. In the M3 panel, 
evolved stars are shown as crosses, and low-amplitude stars are shown as
triangles. See Sect. 3.2.3 for details.  
\label{cacciari.fig4}}
\end{figure}

In Fig. \ref{cacciari.fig4} we compare the A$_V$ amplitude (more abundant 
in literature than A$_B$) vs period distributions for the RRc and RRab 
variables in nine globular clusters with the analogous data in M3.  
The clusters are: three OoII types, i.e. M15, M68 and M9 (data from 
Silbermann \& Smith 1995, Walker 1994, and Clement \& Shelton 1999a, 
respectively); three intermediate types, i.e. IC4499, NGC6934 and NGC1851 
(data from Walker \& Nemec 1996, Kaluzny et al.\ 2001, and Walker 1998, 
respectively); and three OoI types, i.e. NGC3201, M5, and NGC6362 (data from 
Piersimoni et al.\ 2002, Kaluzny et al.\ 2000, and Olech et al.\ 2001, 
respectively). 
From the original datasets we have excluded the stars with indication of 
Blazhko variability or too noisy light curves.  
The values of metallicity shown in Fig. \ref{cacciari.fig4} have been taken 
from KI for all clusters except IC4499, M9 and NGC6934, for which the values 
listed by Harris (1996) have been used. 
In each panel we have reported also the average distributions of the M3 
regular (solid lines) and evolved (dotted lines) RRc and RRab variables, for 
ease of comparison. We note the following: \\
i) OoI and intermediate type clusters show similar distributions to M3 
irrespective of metallicity, including the presence and behaviour of 
evolved stars that in a few cases (e.g. M5, NGC1851) appear to be quite 
abundant. This applies to both RRc and RRab variables.  \\
ii) In OoII clusters the 
distributions of both RRc and RRab stars are again independent of 
metal abundance, and most stars fall on the corresponding distributions of 
the evolved stars in M3. 

Therefore we confirm previous results that there is a unique P-A relation
independent of metallicity for RRab variables in OoI (and intermediate) type 
clusters (Brocato et al.\ 1996; Clement \& Shelton 1999b). However, 
contrarily to previous suggestions of a possibly different metal-dependent 
P-A relation for first overtone pulsators (Clement \& Shelton 1999b), we 
find that there is a unique P-A relation for the RRc as well. 
The same happens in OoII clusters: they too are characterized by their 
own typical P-A relations independent of metallicity, that correspond quite 
closely to the relations of the {\em evolved} RRc and RRab stars in OoI 
clusters. 
This strongly supports the interpretation of the Oosterhoff 
dichotomy as due to evolution away from the ZAHB (cf. Lee et al.\ 1990; 
Clement \& Shelton 1999b).  

\subsubsection{The Blazhko variables}  
We have reported in Fig. \ref{cacciari.fig2} upper and lower panels also the 
{\em known} Blazhko variables (see Table \ref{blafot}), shown as lines 
connecting the 1992 and 1993 CC01 results. 
They do not appear in the middle panel because the Blazhko light 
curves are generally affected by large photometric scatter and the rise 
times are quite uncertain. In general we note that: \\ 
i) they all fall within the RRab group, except one (V44) that could possibly 
belong to the RRc group from the shape of its light curve at minimum 
amplitude, but the photometry is rather scattered; \\
ii) a few of them, when observed at large amplitude, fall on 
the distribution of the 9 longP/overluminous RRab stars that we have 
identified above (cf. Fig. \ref{cacciari.fig3} where they are shown as 
open squares, for completeness). These stars are V3, 14, 24, 35 and 67, 
and their average magnitude (at large amplitude) is $<V>$=15.54$\pm$0.04, 
just like the longP/overluminous RRab stars. \\
We discuss in more detail the nature of all the longP/overluminous 
stars in Sect. 5.2. 
 
Once again we stress that if the RRab stars were considered all together, 
without any knowledge of their Blazhko nature, the scatter of the distribution 
would be large enough to hide completely these sub-groups and their behaviour; 
also the mean relation in the period-amplitude plane would be less well 
defined and likely different in shape and/or zero-point.

\subsection{The Color-Rise Time, Color-Amplitude and Color-Period diagrams} 

\begin{figure}
\plotone{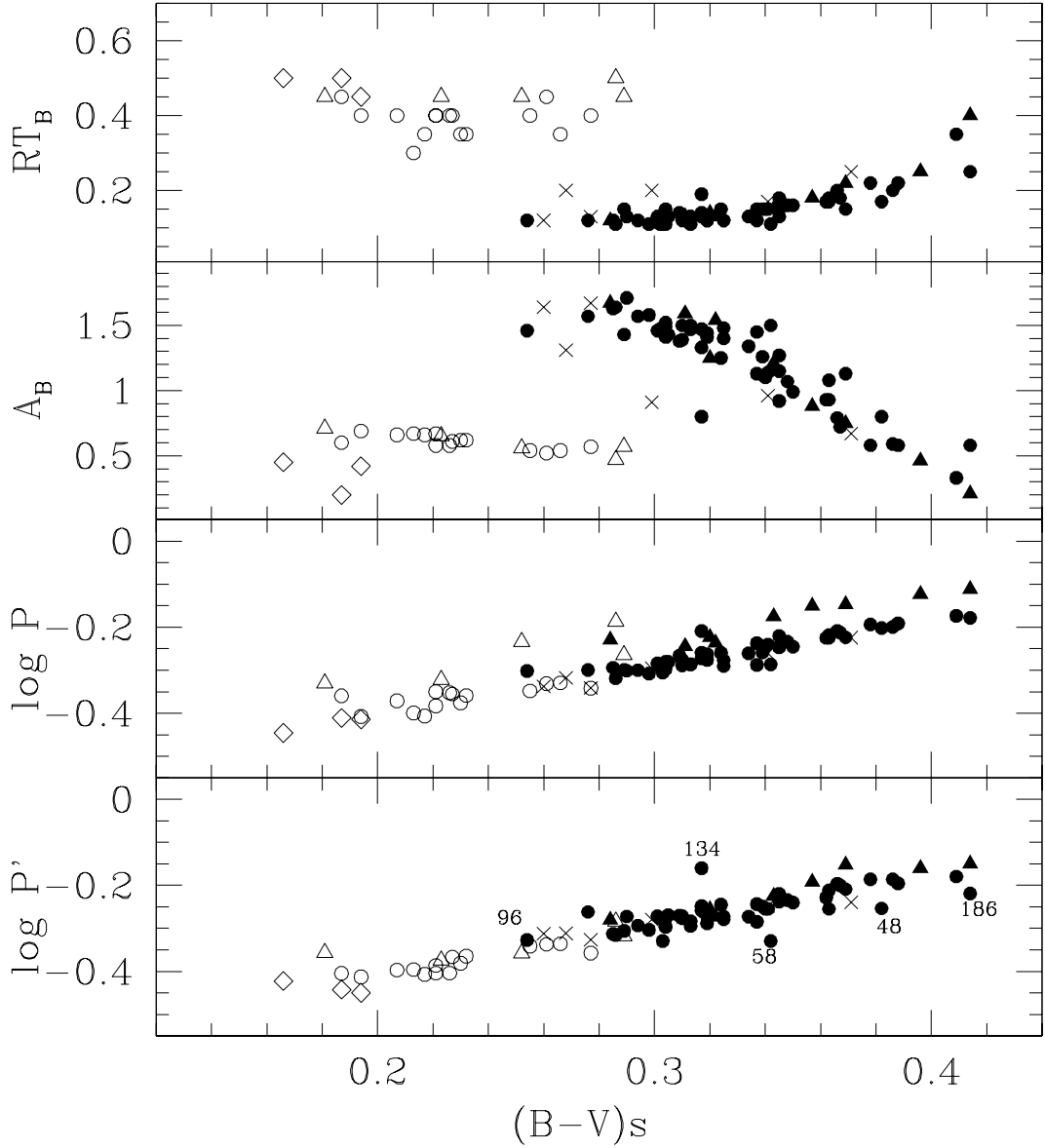}
\caption{Color-rise time, color-amplitude and color-period plots for the 
RRc (open circles), RRab (filled circles), long-period/overluminous 
RRc (open triangles) and RRab (filled triangles), small-amplitude/suspected 
Blazhko RRab (crosses) and short-period small-amplitude RRc stars (diamonds).  
The periods of the RRc variables have been converted to fundamental mode 
periods by adding 0.127 to the $\log\,P$.  
The bottom panel shows the same as the middle panel but for the reduced 
period, defined as $\log\,P' = \log\,P + 0.336(<V>-15.64)$. 
\label{cacciari.fig5}}
\end{figure}

We show in Fig. \ref{cacciari.fig5} the relations between the (B--V)$_S$ color and 
$RT_B$, $A_B$  and $\log\,P$,  for all our RRc and RRab stars. 
The periods of the RRc stars have been fundamentalized by adding 0.127 to the 
$\log\,P$. The bottom panel shows the reduced period 
$\log\,P' = \log\,P + 0.336(<V>-V_{ave})$ that is designed to take 
into account the intrinsic spread in magnitude of the variables and correct 
the periods accordingly (Bingham et al.\ 1984). For $V_{ave}$ we intend the 
mean magnitude of all ``normal'' stars and we use the value 15.64 that 
was derived from the RRab stars in the previous section. As one can see, the 
relation in the bottom panel is indeed tighter. 
In general, the use of the reduced period helps decrease the scatter of the 
period-color relation;  in a few cases, however, it may bring out stars that 
show  normal pulsation characteristics (i.e. periods and amplitudes) but 
unusual photometric properties, e.g. their colors are slightly too red or too 
blue and/or their magnitudes are slightly too bright or too faint. 
In Fig. \ref{cacciari.fig5} we see four possible such stars that slip out of the 
mean relation when the reduced period is used, i.e. V48, V58 and V186, and 
V134. For the first three, the average magnitude is $<V>\sim$15.51$\pm$0.02, 
and we note that V58 and V186 are marked in Table \ref{abfot} as having 
photometric problems. 
Incidentally, these characteristics might be compatible with the presence 
of an undetected faint and redder companion, for example a subgiant 
star at V$\sim$18 and (B--V)$\sim$0.6, of which there is abundance 
in globular clusters. As for V134, it is the faintest star of our sample 
and is unusually blue for its period and amplitude, yet quite normal 
in the period-amplitude plane (cf. Fig. \ref{cacciari.fig3}). 
 
In the color-amplitude  plane  there is a clear correlation between these 
parameters for the RRab stars, whereas the distribution of the RRc stars 
is nearly flat. The three RRc stars with peculiarly small amplitude (V105, 
V178 and V203, shown as diamonds) are clearly distinguishable off the main 
distribution.  In the period-color plane these 
three stars follow the same relation as the main RRc group, whereas  
the longP/overluminous RRc stars stand out from the rest, as expected, but 
fall nicely on the mean relation when using the reduced period that corrects 
for their unusually high luminosity. 

In the RRab group, only two of the low amplitude/suspected Blazhko stars 
(V22 and V54) fall out of the main relation in the  $A_B$ and $RT_B$ vs. 
(B--V)$_S$ planes, but are quite normal in the period-color plane. 
Two more stars, V96 and V134, have unusually blue colors in all planes, 
but they have been marked as having some photometric uncertainties 
(see Table \ref{abfot}). 
Finally, the stars labelled as longP/overluminous look mostly normal in the 
$RT_B$ and $A_B$ vs. (B--V)$_S$ planes, fall out of the mean relation when  
period is involved, but look again normal if we use the reduced period that 
takes into account the effect of (over)luminosity. 
We note that both RRc and RRab variables seem to define quadratic rather 
than linear relations in the $RT_B$ vs. (B-V)$_S$ plane, independently of 
their evolutionary status. 

\begin{figure}
\plotone{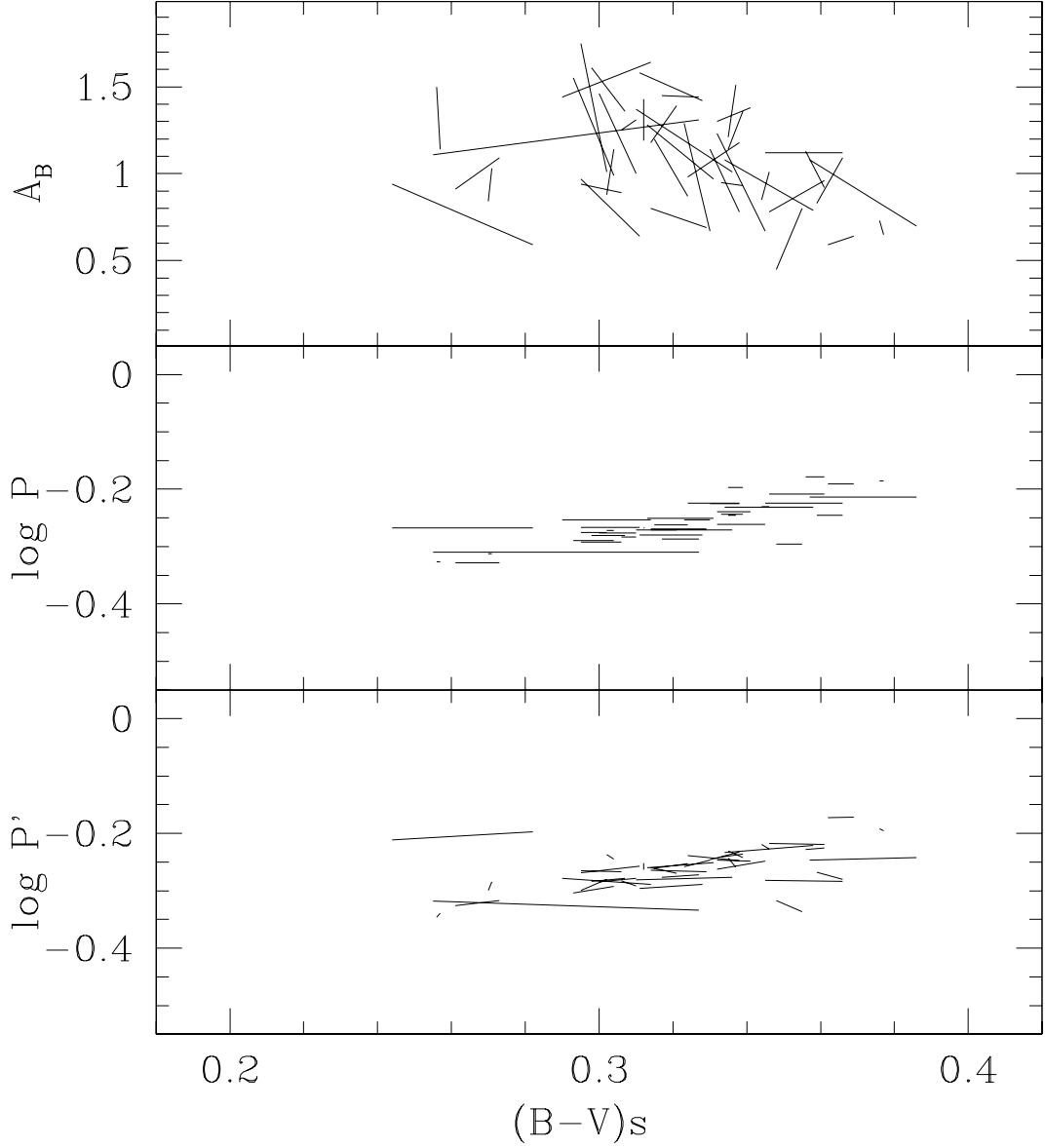}
\caption{Color-amplitude and color-period plots for the Blazhko stars.
The bottom panel shows the same as the middle panel but for the reduced 
period, defined as $\log\,P' = \log\,P + 0.336(<V>-15.64)$. As before, 
the 1992 and 1993 data are considered separately and joined by a line.
\label{cacciari.fig6}}
\end{figure}

Similar plots are shown in Fig. \ref{cacciari.fig6} for Blazhko stars, 
excluding the $RT_B$ data that are not well defined for these stars.

\section{Calibrations for the determination of physical parameters} 

\subsection{Reddening}

An accurate determination of the interstellar reddening is 
essential before we can derive physical parameters, such as temperature 
and luminosity, from observed parameters such as colors and apparent 
magnitudes. 

We consider two approaches to estimate the reddening, using the colors of 
the RRab variables at minimum light $(B-V)_{min}$ that are listed in Table 
\ref{abfot}, and using the mean $(B-V)_S$ colors.    

\noindent $\bullet$ 
Sturch (1966) method uses the (B--V)$_{min}$ color of an RRab star, its period 
and metallicity to derive its reddening E(B--V). 
Among the most recent rediscussions and calibrations of this method are 
Blanco's (1992), based on photometric color indices of field RRab stars 
with known metallicity (via $\Delta S$), and  Walker's(1998), based on  
Sturch's stellar sample and the assumption of zero reddening at the Galactic 
poles. They both find reddening values on average $\sim$0.02 mag larger than 
most other determinations.  
Also Walker (1994) and Walker \& Nemec (1996) find typically 
$\sim$0.02 mag larger values with this method than with methods involving the
color of the red giant branch in the globular clusters M68 and IC 4499.
Therefore, we use the formulation proposed by Walker (1998), 

\noindent $E(B-V) = (B-V)_{min}-0.24P-0.056[Fe/H]-0.356$  \hfill (1) 

\noindent 
where the zero-point has been corrected by --0.02 mag to take this offset into 
account. For the metallicity, the most recent spectroscopic determinations 
are from Kraft et al.\ (1992) who derived [Fe/H]=--1.47 based on high-dispersion 
spectra of a few red giant stars; this result was then confirmed as 
[Fe/H]=--1.50$\pm$0.03 from a new analysis by KI based on Fe $II$ abundances 
of 23 giants. Independently, Sandstrom et al.\ (2001) obtained 
$[Fe/H] \sim -1.22\pm0.12$ from 29 RR Lyrae and 5 red giant stars using low  
resolution spectra, but they note that ``the use of low resolution spectra 
generally causes an overestimate of about 0.25 dex in the derived abundances''.
Therefore we have adopted [Fe/H]=--1.5 for M3. Considering only the RRab 
stars listed in Table \ref{abfot} with good photometry and no evidence of any 
peculiarity, we obtain an average reddening E(B--V)=0.014$\pm$0.012 for M3. 

\noindent $\bullet$ Piersimoni et al.\ (2002) have defined empirical 
period-color-amplitude-(metallicity) relations based on several cluster 
and field RRab variables for which reliable photometry and reddening 
estimates are available. Their relation: 

\noindent $(B-V)_0 = 0.507 - 0.052A_B + 0.223\log\,P + 0.036[Fe/H]$  \hfill (2) 

\noindent allows us to derive the reddening by comparison with the  
average observed color, i.e. (B--V)$_S$.  
Using again the 45 RRab stars in Table \ref{abfot} that show no evidence of 
peculiarity we find an average E(B--V)=--0.001$\pm$0.016.  

Independent estimates, such as those obtained from dust maps (Schlegel et al.\ 
1998), and by comparing the stellar content with the DIRBE/IRAS 100 $\mu$m 
dust emission (Dutra \& Bica 2000), both suggest a value of E(B--V)$\sim$0.01 
for M3. 

A straight average of all these results leads to E(B--V)=0.01, with an r.m.s. 
error of $\sim$0.01 to account for both internal and external uncertainties. 
This is the value generally accepted in all recent studies for M3 and we adopt 
it in the following analysis. 

Considering that for 20 program stars Car98 I-band data are available, 
we could in principle  
estimate the reddening from the (V--I) colors using the relation given 
by Mateo et al.\ (1995), who estimated that the intrinsic (V--I) 
color of RRab variables at minimum light, $(V-I)_{0,min}$, is nearly 
constant with a value of 0.58 $\pm$ 0.03 mag irrespective of metallicity.   
However, the I photometry by Car98 is likely affected by calibration 
problems, as we mentioned in Sect. 2.1, therefore this method could instead  
be used the other way around: from the 20 stars listed in Table \ref{abfot} 
that have I photometry and have no photometric peculiarity we derive an 
average $(V-I)_{min}=0.51\pm0.04$.  
Therefore, as a byproduct of this analysis and a consequence of the adopted 
value of reddening for M3, we find that the correction to apply 
to Car98 I photometry as a calibration offset is $\sim$ --0.083 mag.

\subsection{Calibration in $T_{eff}$ and $m_{bol}$}

A correct determination of temperature is of basic importance for the 
subsequent determination of the stellar physical parameters. 
For this purpose, the reddening must be known as accurately as possible, 
and the best color and color-temperature tranformation equation 
must be used. We feel confident that a reliable value for the reddening is 
available (cf. Sect. 4.1).  We have defined from our data a mean (B--V)$_S$ 
color that is as close as possible to the  static color of the equivalent 
non-pulsating star (cf. Sect. 2.3).  
However,  it has been argued that blue colors may be distorted by shock-induced 
effects in non static atmospheres (see Sect. 2.3), and that infrared 
(e.g. V--K) colors are better temperature indicators (Liu \& Janes 1990; 
CSJ; Cacciari et al.\ 1992, and references therein). 
K photometry is available only for a small number 
of RR Lyrae stars in M3 (29 stars of which 9 RRc and 20 RRab, Longmore et al.\ 
1990, hereafter L90), and we have used these data to test the dependence of 
temperature on the choice of color. 
We did not try to use the (V--I) colors that are available 
for a good number of stars, because we don't think they are accurate or 
reliable enough for this purpose.

\subsubsection{The temperature scales}

For this test we have used six different temperature scales. All of them 
are listed in Table \ref{tecal} except CSJ', which is independent of 
color. We discuss them below in some detail. 

\begin{figure}
\plotone{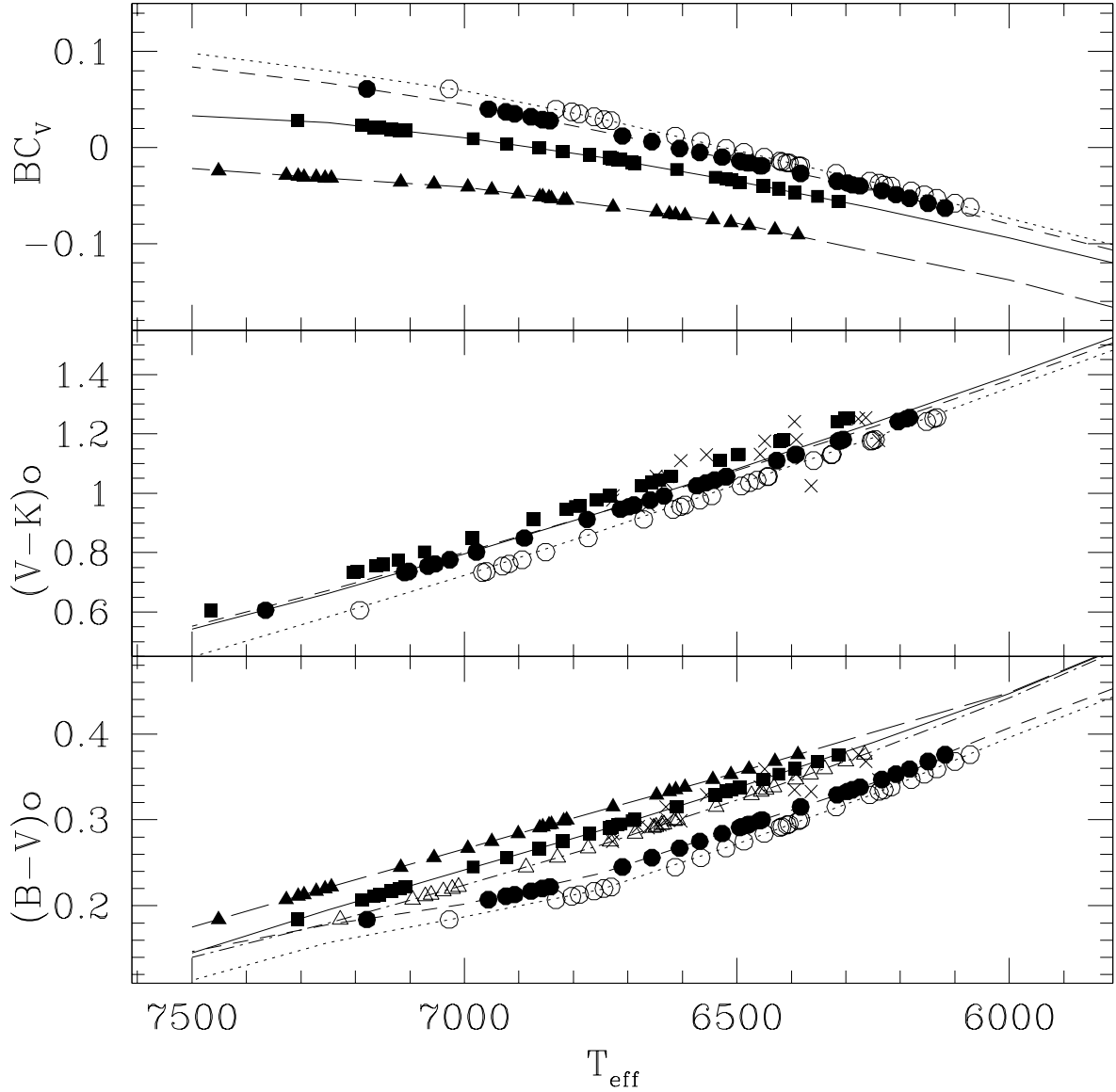}
\caption{Comparison of temperatures as a function of (B--V)$_0$ (lower panel)
and (V--K)$_0$ (middle panel) colors using the calibrations discussed 
in Sect. 4.2.1. Symbols indicate the various calibrations: M98 empirical 
(open circles), M98 theoretical ((filled circles), SF (open triangles), 
C99 (filled squares), SBT (filled triangles), and CSJ (crosses). The upper 
panel shows the $BC_V-T_{eff}$ relations for the calibrations that provide 
them independently. 
\label{cacciari.fig7}}
\end{figure}

\noindent $\bullet$  
The model atmospheres by Castelli (1999, hereafter C99) are based on 
Kurucz models and were calculated with the standard mixing-length 
treatment of convection with no overshooting; we have selected the model 
with metallicity [m/H]=--1.5 and $\alpha$-element enhancement 
$[\alpha/\alpha_{\odot}]$=+0.4, 
and turbulent velocity V$_{turb}$=4 kms$^{-1}$ which seems more appropriate 
for pulsating stars than the usual value of  2 kms$^{-1}$ (we note that  
models with V$_{turb}$=4 kms$^{-1}$ instead of 2 kms$^{-1}$ produce higher 
temperatures  by  $\sim$ 7 and 20 K at (B--V)=0.2 and 0.4, respectively).  
We have adopted for all program stars $\log\,$g=2.75 interpolating linearly 
in the models for $\log\,$g=2.5 and 3.0. We remind the reader that Kurucz' 
(hence C99) models give a solar bolometric 
correction of $-$0.192, therefore the values of BC$_V$  have been corrected 
by adding +0.122 to the model values, as we assume BC$_V({\odot})$=--0.07 
(corresponding to M$_{bol}({\odot})$=4.75) to be consistent with 
Montegriffo et al.\ (1998, hereafter M98) calibration. \\ 
\noindent $\bullet$  The empirical calibration by M98 (M98e) is based on 
Population II giants, namely about 6500 RGB and HB stars in 10 globular 
clusters, that were observed in both optical and near-IR bands. 
This relation is based on and works best for (V--K) colors, but is 
defined also for (B--V) albeit with a lower level of accuracy. \\
\noindent $\bullet$ M98  provide also a theoretical temperature scale (M98t) 
based on Bessell et al.\
(1998) solar metallicity models scaled to lower metallicities by the use of 
C99  models. \\
\noindent $\bullet$ Sandage et al.\ (1999, hereafter SBT) present a new set of 
model atmospheres for temperatures between 5000 and 7500 K. We have considered 
the (B-V) colors and bolometric corrections of the models with [A/H]=--1.5 
(by linear interpolation between the bracketing models at --1.31 and --1.66), 
turbulent velocity V$_{turb}$=5 kms$_{-1}$, and $\log\,g$=2.75 (by linear
interpolation between the bracketing models at 2.25 and 3.0), as shown in 
Fig. \ref{cacciari.fig7}. (V-K) colors are not given by SBT. \\
\noindent $\bullet$ Sekiguchi \& Fukugita (2000, hereafter SF), using 270 
$ISO$ standard stars with accurate estimates of temperature (from IR colors)
and known values of metallicity, gravity and (B--V) color, have derived a 
(B-V) color-temperature relation which they think is the least 
model-dependent. This relation holds for both dwarf and giant stars in the 
range F0-K5 (0.3$\le(B-V)\le$1.5) with metallicity [Fe/H]=--1.5 to +0.3, and  
is parameterized in their eq. (2) that takes into account the 
contributions of (B-V), the gravity and the metallicity. 
No bolometric corrections are given, so we have derived them from C99 
$T_{eff}-BC_V$  relation, because of the similarity of these two
temperature scales. \\
\noindent $\bullet$ Finally, CSJ discussed in detail the problem of the 
temperature determination and proposed a set of equations, of which one 
(their eq. 16): 

\noindent $T_{eff} = 5040/(0.261\log\,P - 0.028A_B + 0.013[Fe/H] + 0.891)$ 
\hfill (3) 

\noindent depends on period, B light curve amplitude and metallicity and is 
independent of color, and
is claimed to give ``the best results in the derivations of equilibrium 
temperatures for RR Lyrae stars''. This parameterization is defined only 
for RRab variables. We have also used this method to derive another estimate 
of temperature for the RRab stars. The values of BC$_V$  that we use along 
with CSJ temperatures have been derived by interpolation in C99
models, because of the similarity of these two temperature scales.

\subsubsection{A test based on the infrared colors}

We have considered the 29 RR Lyrae stars observed in the K-band 
by L90. These data are in the UKIRT photometric system, and 
before proceeding with the application of the above calibrations we must 
ensure that all K values are reported to a homogeneous system. We can do 
that by using 2MASS as an intermediate step and the relations 
between the relevant IR photometric systems given by Carpenter (2001). 
C99 K values are based on Bessell \& Brett (1988) system (cf. Kinman \& 
Castelli 2002), whereas M98 K values are in the ESO system which, according 
to M98 Table 2, is 0.056 mag brighter than Bessell \& Brett in the K band. 
For these systems Carpenter (2001) gives the following relations: 

\noindent $K_{UKIRT} = K_{2MASS} - 0.004(J-K)_{2MASS} - 0.002 $ \hfill (4)

\noindent $K_{BB} = K_{2MASS} + 0.044$  \hfill (5)

\noindent hence we deduce that L90 values of K must be made fainter 
by $\sim$0.047 mag when used with C99 models, and brighter by $\sim$0.009 mag 
when used with M98 calibrations, assuming that (J--K)$\sim$0.25$\pm$0.1 
represents the color range of the instability strip.   

With these corrections to L90 K photometric data, we have  calculated 
the values of temperature $T_{eff}$ from the (V--K) colors,using C99 and M98 
calibrations. The colors were obtained as $<V>$ (taken from Tables 
\ref{cfot} and \ref{abfot}) minus $<K>$ (from L90), and are 
a fairly good approximation of the average colors given the low 
amplitude and nearly sinusoidal shape of the (V--K) curves.  
We assumed a reddening E(B--V)=0.01 and E(V--K)=2.76E(B--V) (Mathis 1999).     
We have calculated $T_{eff}$  also from the (B--V)$_S$  colors listed in 
Tables \ref{cfot} and \ref{abfot}, using all the color-temperature 
calibrations presented above, and using the color-independent relation by 
CSJ.

We show the results of this test in Fig. \ref{cacciari.fig7} and in 
Table \ref{compar} where we present the average values for the RRc, RRab 
and RRc+RRab separately, to retain some information on the trend with 
temperature that is clearly visible in the figure. 

We note the following: \\
i) (B--V)$_S$ colors lead to temperatures that may differ by up to $\sim$450 K 
at (B--V)$_S\sim$0.3, SBT giving the hottest temperatures and M98e giving the 
coolest. This is not surprising. C99 commented on this effect noticing that 
all models give higher temperatures than the empirical relations, generally by 
about 200 K. This discrepancy is reduced by a factor $\sim$2 if (V--K) colors 
are used. By comparison, CSJ values are very similar to SF and they both fall
in the middle range of the considered temperature scales, somewhat cooler 
than C99. \\
ii) Within the same calibration, $T_{eff}$(V--K) are quite similar to 
$T_{eff}$(B--V)  in the C99 calibration, and are instead hotter than 
$T_{eff}$(B--V) by $\sim$100 K in M98. \\
iii) The values of BC$_V$ are quite similar within M98 calibrations, 
the empirical scale giving systematically larger values than the theoretical 
one by $\sim$0.01-0.02 mag.; the C99 scale gives smaller values than M98's by 
$\sim$0.02-0.04 mag, and SBT gives smaller values than C99 by a
further $\sim$0.04 mag.
 
In summary, the temperature calibrations we have considered produce the least 
dispersed values of temperatures when (V--K) colors are used, and we seem to 
be able to define an average temperature scale with an {\em internal} 
uncertainty somewhat smaller than $\pm$100 K if we could use (V--K). 
However, color-temperature scales with (B--V) lead to a dispersion
about twice as large. In general, there may be systematic errors of up to 
200-300 K due to photometric calibrations, transformations and choice of 
temperature scale, and we cannot say which one of these relations is the most 
correct in {\em absolute physical} terms, unless we perform tests and 
comparisons with other parameters derived independently.
 
To this purpose, in the following sections we shall further verify the impact 
of temperature on the determination of physical parameters such as luminosity 
and mass, by confronting the results of pulsation and evolution theories and 
independent observational evidence.

\subsection{The Mass-to-Light Ratio} 

From the pioneering work of van Albada \& Baker (1971) on stellar pulsation, 
it is known that the period of a fundamental mode pulsator 
is related to its mass, luminosity and temperature via the relation: 

\noindent $\log\,P_0=11.50+0.84\log\,L-0.68\log\,M-3.48\log\,T_{eff}$ \hfill (6)

\noindent where $M$ is the mass of the star and $L$ its bolometric luminosity, 
in solar units.  A recent redermination of this relation based on 
non-linear pulsation models by Bono et al.\ (1997) 
includes also some dependence on metallicity (Caputo et al.\ 1998), i.e.:

\noindent $\log\,P_0=11.242+0.841\log\,L-0.679\log\,M-3.410\log\,T_{eff}+
0.007\log\,Z$ \hfill (7)



\noindent If we consider that [$\alpha$/Fe]$\sim$0.3 for M3 (Kraft et al.\ 1993, 
1995) would mimic a total metallicity content [m/H]$\sim-$1.3 
(cf. Salaris et al.\ 1993), then $\log\,Z$=--3.06. 
Therefore the effective temperatures and periods of the variables can be used 
to derive a mass-luminosity parameter $A$ for the fundamental pulsators defined  
as:

\noindent $A=0.81\log\,M-\log\,L=13.353-1.19\log\,P_0-4.058\log\,T_{eff}$ 
\hfill (8)
 


\noindent This definition of the $A$ parameter can be used also for first 
overtone 
pulsators  provided their periods are fundamentalized by adding 
0.127 to their $\log\,P_1$. 
We note that Caputo et al.\ (1998) give a separate pulsation relation for 
the first overtone pulsators, which yields essentially the same results. 

We have applied eq. (8) to derive the $A$ parameter for the 29 test 
RR Lyrae stars that were considered in the previous section, using 
the various estimates of temperature to evaluate their impact on the 
mass or luminosity determination. The values of  $A$  are 
listed in Table \ref{compar}. Incidentally, we note that a further set of 
pulsation relations has become recently available (Marconi et al.\ 2003).
We have verified that they produce systematically larger values of the $A$ 
parameter by $\sim$ 0.014 (RRc stars) and 0.011 (RRab stars) respectively, 
that translate into larger masses by 0.01-0.02 M$_{\odot}$ at fixed 
luminosity, or fainter magnitudes by 0.01-0.02 mag at fixed mass. 
These differences are well below the errors of these estimates and do not 
affect significantly  the following analysis and considerations.

\subsubsection{The luminosity of the RR Lyrae stars assuming a fixed mass}

From the $A$  parameters derived in the previous section we can 
estimate the luminosity of the RR Lyrae stars if we know their mass. 
Masses can be obtained in two independent ways: adopting the values of the 
stellar evolution theory for HB stars, that usually range from 0.65 to 
0.75 $M_{\odot}$, or from the stellar pulsation theory applied to 
double-mode pulsators (RRd). 
The most recent analysis and discussion of 8 RRd stars in M3 by Clementini 
et al.\ (2004) shows an unusually large dispersion in mass for these stars. 
Whether this reflects a similarly large mass dispersion for all HB (hence 
RR Lyrae) stars is not clear. We assume we can consider a constant mass for 
these stars, and for this we take the weighted average of the 
mass determinations for these 8 RRd stars, i.e. 0.74$\pm$0.06 $M_{\odot}$. 
We note, however, that masses of RRd stars are quite uncertain as they depend 
strongly on modelling (in particular on the adopted metallicity scale), and on 
the accuracy of the period determinations (cf. Bragaglia et al.\ 2001). 
On the other hand, the most recent models of HB stellar evolution (e.g. 
Sweigart 1997; Marconi et al.\ 2003) would rather favour a value around 
0.67-0.69 $M_{\odot}$,  so we consider 0.68 $M_{\odot}$ as the evolutionary 
mass of RR Lyrae stars in the following.

Using these values for the stellar mass and the values of 
$A=0.81\log\,M-\log\,L$ (r.m.s. error $\pm$0.03) listed in Table \ref{compar} 
we then derive  the corresponding values of absolute magnitude M$_V$ that 
are also listed in Table \ref{compar}, assuming $M_{bol}(\odot)$=4.75.  
We see that these values vary by up to nearly 0.3 mag, from 0.42 to 0.70,  
are very dependent on the temperature calibrations and quite sensitive 
as well to the adopted mass. Also the choice of color can make a difference, 
in particular the M98 calibrations do not produce consistent results from 
(B--V) vs. (V--K), whereas the discrepancy is much smaller with the C99 
calibration. 

How accurately and precisely do we know the masses of the RR Lyrae stars, 
to start with? This quantity is still quite uncertain, and we have negligible
prospects to improve our knowledge by measuring any masses directly. 
We are more likely to improve our distance estimates to clusters in the 
near future, so M$_V$ will become increasingly well known. Therefore we turn 
the problem around and use the mass-to-light parameter A to estimate the 
mass at fixed luminosity.

\subsubsection{The mass of the RR Lyrae stars assuming a fixed luminosity}

If we assume that globular cluster and field RR Lyrae stars share 
the same characteristics (Catelan 1998; Carretta et al.\ 2000), 
we may use the results by Cacciari \& Clementini (2003) who 
estimated from several independent methods the average absolute 
magnitude for the RR Lyrae stars with [Fe/H]=--1.5 as 
$<M_V>$=0.59$\pm$0.03 mag. This is in agreement with the
most recent synthetic HB models by Catelan et al.\ (2004) that would 
predict $<M_V>\sim$0.6 at $\log\, Z$=--3.06 (corresponding to 
[Fe/H]=--1.5 with $\alpha$-element enrichment +0.3).
On the other hand, the accurate study of RR Lyrae stars in the LMC by 
Clementini et al.\ (2003) and the pulsational distance modulus of 
15.07$\pm$0.05 mag for M3 estimated by Marconi et al.\ (2003) would favour 
a brighter value around 0.54-0.55 mag. We therefore consider 
that values of $<M_V>$ in the range 0.54-0.59 mag are quite reasonable 
based on independent empirical and/or  theoretical considerations. 
These values, inserted in  eq. (8), lead to the values of mass listed in 
Table \ref{compar}. Typical error of these determinations is 
$\pm$0.05 M$_{\odot}$. 

Again, we see that the values of mass range from about 0.6 to nearly 
0.8 M$_{\odot}$ between calibrations, but vary by less than 
0.05 M$_{\odot}$ within each color/calibration.  
However, the estimates in the restricted range $\sim$0.68-0.74 M$_{\odot}$, 
that we regarded as ``plausible'' in the previous section, are not so many. 
Limiting for simplicity to the RRab variables, that are more numerous 
hence better representative of the entire population, 
only the M98 scales lead to acceptable results using (V--K) colors. 
With (B--V) colors, acceptable results come from M98 theoretical 
and SF (which gives nearly identical results to CSJ).  

To summarize, in order to optimally exploit our large and accurate database 
we need to use (B--V) colors, since (V--K) colors are available only for few 
stars, and the color-independent scale of CSJ is only applicable to RRab 
variables. 
On this basis, there are two temperature scales that may provide plausible 
estimates  of both mass and luminosity, i) M98 theoretical, leading to fainter 
and more massive stars (in agreement with the most recent results on the
mass of RRd stars, with theoretical HB models by Catelan et al.\ 2004, and 
with the average of several different estimates of absolute luminosity for 
RR Lyrae stars), and ii) SF, leading to slightly brighter and less massive 
stars (in agreement with the most recent estimate of distance to the LMC and 
with stellar evolution and pulsation models). However, the M98 
calibrations are more accurate and reliable when used with infrared colors 
(cf. Sect. 4.2.1), that would rather support the brighter and less massive 
solution. Therefore we assume for M3 the distance modulus (m--M)$_0$=15.07 
and adopt the SF calibration as a working hypothesis 
for our subsequent analysis, keeping in mind that a somewhat cooler 
temperature scale (e.g. by $\sim$150 K) or a shorter distance modulus (e.g. 
by $\sim$0.05 mag) might be also acceptable.

\begin{figure}
\plotone{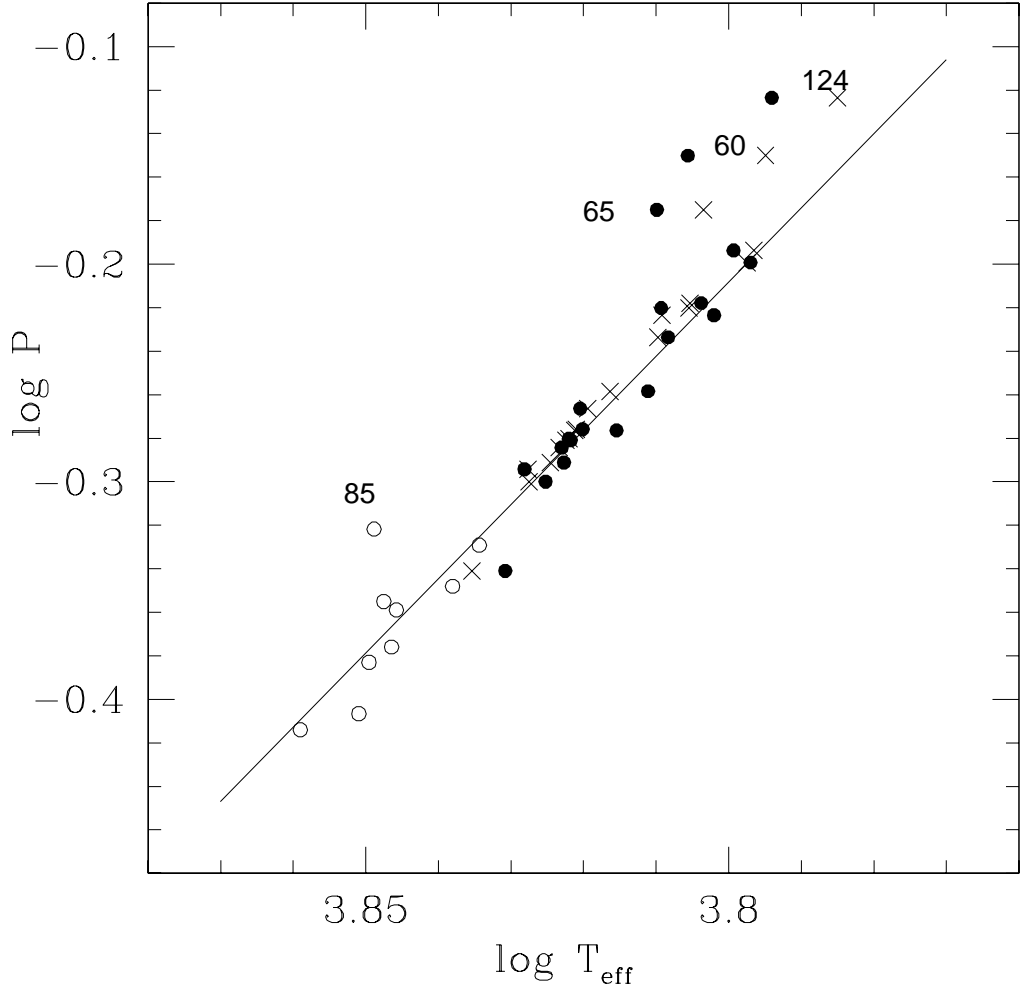}
\caption{Period vs temperature for the 29 test stars discussed in Sect. 4.
RRc are shown as open circles, RRab stars as filled circles. 
The periods of the RRc stars have been fundamentalized by adding 0.127 to 
their $\log\,P$. The temperatures have been obtained from the (B--V)$_S$ 
colors and the SF temperature scale. The line indicates the best fit to the 
data using a slope of --3.41 (cf. eq. 7), the zero-point corresponds to 
$A$=--1.82. The four labelled stars are evolved off the ZAHB. 
The crosses indicate the same RRab stars, where the temperatures have been 
derived from the CSJ scale (eq. 3), for comparison. The mean relation 
represents well also these data, and the scatter is significantly reduced. 
The three evolved RRab stars are still offset from the mean relation, but 
the shift in period ($\Delta \log\,P$) at fixed temperature decreases from 
$\sim$0.069 (with the SF scale) to $\sim$0.043 (with the CSJ scale).    
\label{cacciari.fig8}}
\end{figure}

In Fig. \ref{cacciari.fig8} we  show the fundamentalized period $P_0$ vs 
$T_{eff}(B-V)$ for our 29 test stars. The line shown in the plot indicates 
the best fit to the data using a slope of --3.41 according to eq. (7). 
The corresponding  zero-point yields a value $A$=--1.82$\pm$0.03. 
The longP/overluminous stars V85 (RRc) and V60, V65 and V124 (RRab)  
stand out clearly in this plot. 
For comparison, we show also the $T_{eff}$ 
values obtained from the CSJ temperature scale (eq. 3) for the same RRab 
stars (indicated as crosses). We see that the SF and CSJ scales are very 
similar because both sets of temperatures fit the same 
$\log\,P-\log\,T_{eff}$ relation, the CSJ determinations with a
significantly reduced scatter. 
The three evolved RRab stars still fall clearly 
off the main relation, but by a smaller amount, and this leads to a smaller 
value of the period shift at fixed temperature, i.e. $\Delta \log\,P$ 
decreases from $\sim$0.069 (with the SF temperatures) to $\sim$0.043 
(with the CSJ temperatures).    
This shows the great potential of the CSJ reddening-independent temperature 
parameterization and its application to those cases where reddening can be a 
problem or the data are not sufficiently accurate for a good definition of 
the mean color.   

\section{The physical parameters of our program RR Lyrae stars} 

We have applied SF color-temperature calibration to all our RRc,  RRab 
and Blazhko stars using the (B--V)$_S$ colors listed in 
Tables \ref{cfot}, \ref{abfot} and \ref{blafot}. 
For the Blazhko stars we have used the average value of the 1992 and 1993 
CC01 photometric data. We have calculated the corresponding values of 
temperature, hence bolometric corrections borrowing the C99 
$T_{eff}-BC_V$ scale.  Bolometric magnitudes and luminosities were then 
obtained from the values of $<V>_0$ and (m--M)$_0$=15.07, and the $A$ 
parameter and the mass were estimated from eq. (8). The results  are 
listed in Table \ref{phyabc} for the RRc and RRab stars, 
and in Table \ref{phybla} for the Blazhko stars. 
For the sake of completeness, we have calculated the same physical parameters
using the CSJ temperature calibration (for RRab stars only) expressed in 
eq. (3), and we compare the results in the following sectios whenever relevant.  
 

Typical errors for the above parameters of each individual star are 
$\Delta T_e$=$\pm$100 K,  $\Delta BC_V$=$\pm$0.02 mag, 
$\Delta \log\,L$=$\pm$0.03, $\Delta M_V$=$\pm$0.07 mag, $\Delta A$=$\pm$0.03, 
$\Delta M/M_{\odot}$=$\pm$0.05  and $\Delta \log\,g$=$\pm$0.10.

\subsection{Comparison with evolution and pulsation models}

With our database and the adopted temperature calibration we may test 
recent theoretical models of HB evolution and RR Lyrae pulsation, within 
the limits of the respective uncertainties.  
We show in Fig. \ref{cacciari.fig9} how periods and the physical parameters we have 
derived in the previous sections behave as a function of temperature, for all 
our program RRc and RRab variables. The Blazhko stars are not shown to avoid 
confusion, but they behave like the RRab variables. The results obtained 
from the CSJ temperature calibration are shown in Fig. \ref{cacciari.fig10}.  
Two recent studies of the evolutionary and pulsational characteristics of 
M3 RR Lyrae variables (Marconi et al.\ 2003; Catelan 2004) provide detailed 
theoretical reference frames, for comparison. 
The results proposed in those papers are compatible with the considerations 
presented below.

\subsubsection{$\log\,P$ vs $\log\,T_{eff}$}

First, we compare our results with the basic requirements of the pulsation 
theory. The preliminary test performed on a subset of 29 stars using  
(V--K) colors (cf. Sect. 4.3) produced a  $\log\,P$--$\log\,T_{eff}$ 
relation that is shown in Fig. \ref{cacciari.fig8}. This relation, 
reported in Fig. \ref{cacciari.fig9} lower panel for ease of comparison, 
represents well also the main body of the RRc and RRab stars whose 
temperatures have been derived from (B--V) colors.  
The stars with unusually long periods, that were noticed in the 
period-amplitude and period-color diagrams (see Sect. 3.2 
and 3.3), stand out clearly in this diagram as well, as expected. 
The same relation is defined by the CSJ temperatures, with a somewhat smaller 
dispersion, as one can see in Fig. \ref{cacciari.fig10}. The evolved stars stand out 
of the main relation, but with a smaller offset corresponding to a smaller 
period shift at fixed temperature. The temperature range defined by the CSJ 
scale is very nearly the same as that defined by the SF scale if one 
excludes the coolest evolved star V202. 

\subsubsection{$\log\,L$ vs $\log\,T_{eff}$}

This is perhaps the most critical diagram because is a place 
where we can in fact test the correctness and accuracy of our own  
calibrations. To perform this test, we have derived for each star the offset 
in $\log\,L$ with respect to the reference ZAHB level we have estimated at 
$\log\,L\sim$1.66 at mid temperature range ($\log\,T_{eff}$=3.83), and 
compared it with the corresponding offset in V magnitude with respect to 
the observed ZAHB level that we have identified at V=15.72 (cf. Sect. 3.1). 
We show the diagram of the $\Delta \log\,L$ vs. $\Delta V$ offsets in Fig. 
\ref{cacciari.fig11}, for all the RRc and RRab stars listed in Tables \ref{cfot} and 
\ref{abfot}.    
We see that the offsets are well represented by a relation of slope 1, and 
fall, with no exceptions, within $\pm$0.07 mag of this relation, which is 
the typical error we have estimated for the luminosity.  The same result, 
with a somewhat smaller scatter, is obtained by using the CSJ temperatures 
(but we note that this reduced scatter is partly due to the missing RRc 
stars). 
Therefore, we deduce that our calibration is reliable and accurate, 
within the uncertainties of these estimates, 
and we proceed with a more detailed comparison with stellar evolution 
models. 

We compare the luminosities and temperatures we have derived with three sets 
of theoretical evolutionary models for the ZAHB phase, i.e. Sweigart (1997, 
solar scaled [Fe/H]=--1.6, main sequence helium 
abundance Y=0.23, no helium mixing during the RGB phase), VandenBerg 
et al.\ (2000, [Fe/H]=--1.54, [$\alpha$/Fe]=0.3) and Straniero et al.\  
(1997, solar scaled [Fe/H]=--1.63). 
Within the  uncertainties the models are all quite similar, and our 
stars are fully compatible with them. The best match is given by the 
Sweigart ZAHB that practically coincides with the lower envelope 
of our distribution ($\log\,L$=1.666 at $\log\, T_{eff}$=3.83), whereas
VandenBerg et al.'s is fainter by $\sim$ 0.02 mag, and Straniero et al.'s
is brighter by about the same amount. 
The evolved stars that were labelled as longP/overluminous in Sect. 3 stand 
clearly out of the main relation. 

We also compare this distribution with the theoretical limits of the 
instability strip calculated by Bono et al.\ (1995) for a helium abundance 
Y=0.24 and two values of the HB stellar mass, 0.65 and 0.75 M$_{\odot}$. 
We see that our instability strip is systematically hotter by $\sim$150 K. 
The temperature of the blue edge of the instability strip, taken as the 
temperature of the second bluest RRc star in Tab. \ref{phyabc}, 
is $\sim$7300 K, and the width of the instability strip is 
$\Delta \log\,T_{eff}$=0.074 (see Marconi et al.\ 2003, and Catelan 2004, for 
recent discussion on the instability strip edges). 
As for the detailed distribution within the strip, 
Bono et al.\ (1995) assumed  M=0.65 M$_{\odot}$ as appropriate for M3 and 
concluded that the zone between the fundamental blue edge (FBE) and the first 
overtone red edge (FORE), where both pulsation modes are possible, 
is populated mostly by RRab stars, and, hence, the direction 
of evolution on/near the ZAHB is mainly blueward. This would be in agreement 
with Lee et al.\ (1990) evolutionary models, and with the interpretation of 
the Oosterhoff dichotomy as mostly due to the hysteresis mechanism 
in the pulsation modes. However, we don't quite see this effect with our data: 
the FBE-FORE zone appears to be populated by a nearly equal number of RRc and 
RRab stars in the M=0.65 M$_{\odot}$ strip, and only at M=0.75 M$_{\odot}$ 
the RRab stars outnumber the RRc. In a likely intermediate solution with  
M$\sim$0.7 M$_{\odot}$ the FBE-FORE zone should still be 
populated by a non-negligible number of RRc stars.  These considerations hold 
also with the temperatures and luminosities derived from the CSJ calibration, 
and would be even more 
important had we used the cooler temperature scale by M98, hence casting some 
doubts on the hysteresis mechanism as the only or most important 
way to explain the Oosterhoff dichotomy (cf. Sect. 3.2.3).  

If period changes and mode switching 
can be taken as indicative of direction of evolution, our result 
is confirmed by the study of period changes by CC01 who find a nearly 
equal number of RR Lyrae stars near the ZAHB with decreasing and increasing 
periods. Also, four double-mode pulsators  have been found 
switching pulsation mode during the last few years: of these, 
three have switched from fundamental to first overtone mode, 
i.e. V79 (Clement et al.\ 1997; Clement \& Shelton 1999b), V166 (Corwin et al.\ 
1999) and V200 (Clementini et al.\ 2004), whereas one (V251) has switched
from first overtone to fundamental mode (Clementini et al.\ 2004), 
suggesting that both redward and blueward evolution can occur among the HB 
stars in M3.  We also point out that CC01 noted several stars as having 
strongly variable periods over the last $\sim$ 30-50 years, among them 
the three longP/overluminous RRc stars V70, V129 and  V170. 
These period variations are too strong to be 
ascribed to a normal rate of evolution, and rather suggest irregularities 
in the pulsation (possibly a prelude to mode switching?): 
V70 and V129 have increasing periods, and V170 has a decreasing period. 

\subsubsection{$Mass$ vs $\log\,T_{eff}$}

In this diagram we see that the values of mass we have derived from the $A$
parameter and the luminosity follow quite well the theoretical 
trend with temperature, with only few stars deviating from the mean 
distribution by more than $\pm$0.1 M$_{\odot}$. The scatter of this relation 
is further reduced by the use of the CSJ temperatures 
(cf. Fig. \ref{cacciari.fig10}). 
From the present data listed in Tables \ref{phyabc} and \ref{phybla} the 
average values of mass for the regular RRab and RRc stars with 
no photometric anomaly are $<M>$=0.71$\pm$0.03  ($<M>$=0.70$\pm$0.05 from 
CSJ temperatures) and 0.70$\pm$0.05 M$_{\odot}$, respectively. 
The average mass of the Blazhko stars is $<M>$=0.70$\pm$0.08, i.e. 
identical to the regular stars. 
These estimates compare very well with the mass values of ZAHB stars in 
Sweigart and Straniero et al.\ evolutionary models (i.e. 0.68 and 0.69 
M$_{\odot}$, respectively), but are somewhat larger than the mass of 
VandenBerg et al.'s model,  0.64 M$_{\odot}$. This is due to the 
different temperature scale used by VandenBerg et al.\ (2000), which is 
slightly steeper than SF calibration and is hotter by $\sim$100 K at the 
reference mid range temperature $\log\,T_{eff}$=3.83. 
Such a temperature difference leads to smaller values of the $A$ parameter
by $\sim$0.025, that combined with the slightly smaller luminosities 
(by $\sim$0.008 in the $\log\,$) lead to smaller values of 
the mass by $\sim$9\%, i.e. $\sim$0.06 M$_{\odot}$.

\subsubsection{$\log\,g$ vs $\log\,T_{eff}$}
 
Once the values of temperature, luminosity and mass are known, the gravity 
can be derived from the equation of the stellar structure  (cf. eq. 21). 
In this diagram we see that the values of gravity we have derived follow 
the same trend with temperature as the theoretical predictions, with little 
scatter. This of course mirrors the behaviour of mass, as discussed in the 
previous section. 

From the data listed in Table \ref{phyabc} the average values of 
gravity for the RRc and RRab stars are $<\log\,g>$=2.94$\pm$0.04 and 
2.81$\pm$0.04 (2.80$\pm$0.04 from the CSJ temperature scale), respectively. 
The corresponding values at $\log\,T_{eff}$=3.83 are $\sim$2.86 in 
VandenBerg et al.'s ZAHB model, and 2.88 in the two other models.

\subsection{On the evolutionary status of the RR Lyrae variables and 
the nature of the longP/overluminous stars} 

In order to investigate in some detail the nature of those 19 (5 RRc, 9
RRab and 5 Blazhko) stars that were identified in Sect. 3.1 as 
longP/overluminous we have plotted again in Fig. \ref{cacciari.fig12} the 
values of $\log\,L$ vs $\log\,T_{eff}$ , and for comparison the 
ZAHB models calculated by VandenBerg et al.\ (2000)  and by 
Sweigart (1997) that we have presented in Sect. 5.1.2. 
Here we show some additional models by Sweigart (1997), in particular the 
ZAHB corresponding to helium mixing DX=0.05 during the RGB phase, and 
three evolutionary tracks for no helium mixing  and the 
Reimers (1975) mass loss efficiency parameter $\eta$=0.446, 0.378 and 0.300. 
The helium mixing parameter DX measures the depth into which the mixing 
currents are assumed to penetrate the hydrogen shell: all of the helium 
produced exterior to that point is mixed into the envelope. 
As an example, DX=0.05 increases the envelope helium abundance at the tip of 
the RGB by about 0.03 dex with respect to the no-mixing case DX=0.0 
(cf. Sweigart 1997 for more details).  
We notice the following: 

i) The lower envelope of the RRc and RRab star distribution is very well 
represented by the Sweigart ZAHB with DX=0.0.  Within the errors of these 
determinations, also the VandenBerg et al.\ ZAHB, which is $\sim$0.02 mag 
fainter, may provide an acceptable match. Four RRab stars (V76, V77, V134 and 
V197) fall below the ZAHB: they all have been noted for having somewhat 
incomplete and/or noisy light curves. 

ii) If we consider that an intrinsic thickness of $\sim$0.1 mag of the 
stellar distribution is normal (see also Fig. \ref{cacciari.fig1}), as 
it corresponds to that part of the HB evolutionary phase where the stars 
spend most of their HB lifetime, there is however a good number of stars that 
are brighter than this value. 
For these we can think of three possible explanations: \\
1. The most obvious (and least interesting) suggestion is that the photometry 
for some of these overluminous stars may be contaminated by the presence of a 
companion. Although this phenomenon is likely to be rather unfrequent,  
Table \ref{abfot} notes that 3 of the 
4 brightest RRab stars (V48, V58, V146 and V186, all with 
$\log\,L \gtrsim$1.74) may have companions making them appear to be too bright. 
Further, all four stars do not have unusually long periods for their colors 
or temperatures, indicating they have higher gravities and lower 
luminosities. \\
2. The overluminous stars could be stars that have undergone some degree of 
mixing during the RGB phase, and have therefore a slightly higher abundance of 
helium in the atmosphere which makes them brighter (Sweigart 1997). 
This is compatible with observational spectroscopic evidence, e.g. enhanced 
C-isotope ratios and lithium abundance (Pilachowski \& Sneden 2001), and the 
CN-CH anticorrelation (Lee 1999; Smith et al.\ 1996) among M3 red giants that 
suggests the presence of mixing and dredge-up of processed material (including 
helium) in the atmosphere of these stars. We have plotted for comparison the 
Sweigart (1997) ZAHB model for a mixing value DX=0.05 and we see that all our 
stars fall below this ZAHB, except the three brightest RRc stars. 
The explanation based on the mixing hypothesis can only be tested with a 
high-resolution abundance analysis of these stars, that should possibly be 
extended to include the entire HB from red to blue for a more accurate and 
conclusive analysis of this issue.  Incidentally, high resolution spectra 
can also be used to explore line broadening (provided the exposure times 
are short enough to avoid velocity smearing and the observations are taken 
at carefully selected phases). In this respect we note that Carney 
et al.\ (2003) detected line broadening, that was interpreted as a sign of 
rotation, among luminous RGB and red HB stars in the field, and a similar 
behavior was found among luminous red giants in M3  by Carney et al. (2004). 
Rotation may contribute to either mixing or mass loss, and therefore 
have an influence on HB evolution.   \\
3.  Alternatively, or in addition to the previous explanation, the overluminous 
stars could be in a more advanced stage of evolution off the ZAHB than the 
main body of the RR Lyrae variables. 
For comparison, we have plotted three evolutionary tracks 
from Sweigart (1997) models, corresponding to zero mixing and different 
values of the Reimers mass loss parameter, i.e. $\eta$=0.446, 0.378 and 0.300
(note that the entire ZAHB is described by values of $\eta$ from 0.00 to 
0.74). We have selected these three values because they either start and 
evolve mostly within the instability strip ($\eta$=0.300), or they start 
hotter than the instability strip and evolve across it at a plausible 
luminosity ($\eta$=0.378 and 0.446). Smaller values of $\eta$ start 
cooler than the strip and don't enter it, and larger values of $\eta$ would 
cross the strip at too high luminosity levels and at a very fast evolutionary 
rate. We can see that basically all stars fall on or near the tracks with 
$\eta$ values between 0.300 and 0.378, at various stages of evolution off the 
ZAHB, and some evolved stars would be consistent with higher values of mass 
loss ($\eta \sim$0.446).
In this respect, we don't agree with the conclusion of Jurcsik et al.\ (2003)  
that  ``... in M3, on the average, RRc stars are already in a later 
phase of their HB evolution than the RRab variables''. The difference between 
these two groups could simply be due to stochastically different mass loss 
that causes the less massive stars to populate the first-overtone (bluer) part 
of the instability strip. This would apply to all RRc stars, not only to the 
overluminous ones, in agreement with the fact that the average mass of the RRc
stars seems to be about 0.02 M$_{\odot}$  smaller than the mass of the RRab 
stars (cf. Sect. 5.1.3), although this difference is indeed smaller than the 
r.m.s. errors of these determinations. 

To test the plausibility of this explanation we have examined the HB lifetimes 
at various luminosity levels off the ZAHB using the $\eta$=0.300 track. 
We see that the stars spend $\sim$2/3 of their total HB lifetime (i.e. 
$\sim$65 My) between the ZAHB luminosity level and the brighter magnitudes 
within 0.1 mag of the ZAHB (i.e. in a luminosity interval 
$\Delta \log\,L \sim$0.04), progressively accelerating 
their evolution as they become brighter (i.e. spending 5\% of the total 
lifetime, and then 3.5\%, 3\% and 2\% over the following 0.05 mag steps). 
Nearly 80\% of the total HB lifetime is spent between the ZAHB and the 
0.3 mag brighter level.  

In order to estimate the number of stars in the HB evolutionary 
phase we can use the {\em fuel consumption equation} by Renzini 
\& Fusi Pecci (1988): 

\noindent $N_j = B(t)L_Tt_j$ \hfill (9) 

\noindent where $N_j$ is the number of stars predicted by the stellar evolution 
theory in a given post-main-sequence evolutionary phase, $t_j$ is the lifetime
of that phase in yr, $L_T$ is the total luminosity (in solar units) of the 
stellar system and $B(t)$ is a specific evolutionary flux that is a function 
of age and can be taken as 2.15~10$^{-11}$ for a system as old as M3.  
For $t_{HB}$=10$^8$ yr and $L_T=3.4~10^5 L_{\odot}$ for M3 (Harris 1996), 
$N_{HB}$ turns out to be about 730 stars. 
The morphology of the HB in M3 has been studied in detail with stellar counts 
by Ferraro et al.\ (1997), and from their results we can expect that about  
$\le$ 40\% of all the HB stars evolve and can be detected as RR Lyrae 
variables, e.g. something like 290 stars. This is of course just a rough  
estimate, but quite consistent with the observational evidence (cf. 
Clement et al.\ 2001). 
Of these stars, about 65\% (i.e. 190 stars) are expected to populate the zone 
between the ZAHB and the 0.1 mag brighter level, and about 5\% (i.e. 15 stars) 
the brighter 0.05 magnitude interval, reaching $\log\,L \sim$1.75. 
 
These estimates agree  very well  with the observational data and 
indicate that either one of the above explanations, or more likely a 
combination of both, can account for the presence of the longP/overluminous 
RR Lyraes that have been detected in M3. 

To summarize on the evolutionary status of our target stars, we can compare 
again Fig. \ref{cacciari.fig1} and Fig. \ref{cacciari.fig12} and note that: \\ 
i) the main body of the RRc, RRab (and Blazhko) stars has a luminosity 
distribution with a FWHM $\sim$ 0.1 mag and a range $\sim$ 0.2 mag, 
whose faint end corresponds to the ZAHB at V=15.72. 
These stars are compatible with being in the first 65\% of their HB life. \\
ii) There is a significant number of brighter stars distributed 
more loosely around V$\sim$15.52 and few outliers up to 0.2 mag brighter. 
Most of these stars have longer periods than the stars of similar amplitude 
(temperature), and are compatible with being on a more evolved stage of their 
HB life. Some of these unusually bright stars could have enhanced helium in 
their atmospheres due to extra-mixing during the RGB phase. For some, the 
presence of a companion affecting the photometric data cannot be excluded. \\
iii) Although the evidence is marginally significant, the RRc stars might be 
on average slightly less massive than the RRab stars. There is no evidence 
that they are more evolved.

\section{The Fourier analysis of the light curves}

The description of pulsating variable light curves using Fourier series 
started with Schaltenbrand \& Tammann (1971) on Cepheids, later  
followed by Simon \& Lee (1981), Simon \& Teays (1982) and Simon (1988) on 
RR Lyrae stars. Then, in a series of papers (Simon \& Clement 1993, 
hereafter SC93; Kov\`acs \& Zsoldos 1995; Jurcsik \& Kov\`acs 1995, 1996; 
Kov\`acs \& Jurcsik 1996, 1997; Kov\`acs \& Kanbur 1998; Kov\`acs \& Walker 
2001) it was shown that 
appropriate combinations of terms of a Fourier representation of RR Lyrae 
light curves, along with the periods, correlate with intrinsic parameters of 
the stars such as metallicity, mass, luminosity and colors. 
The most recent analysis of Fourier parameters and their physical meaning 
in the study of RR Lyrae stars is given by Sandage (2004a). 

The Fourier analysis may offer tremendous potential to the study of these stars
(e.g. their absolute magnitude hence distance, intrinsic colors hence reddening 
and temperature, mass and metal abundance), especially since large collections 
of good light curves are becoming available, even for distant stars, thanks to 
the photometric surveys dedicated to gravitational lensing events. 
Therefore, it is important to explore the reliability of this type of analysis 
in more detail, and our M3 data are very useful tests of the several 
calibrations between Fourier coefficients and physical parameters. 
We anticipate our conclusions, namely that all the physical parameters derived 
from Fourier Transform coefficients are affected by some (systematic) 
inaccuracy that in a few instances can be rather serious and make them 
unreliable. In general, more work is needed to reanalyse and recalibrate this  
technique before it can be used with confidence. In the following sections 
we discuss in detail how we have tested the Fourier results and reached our 
conclusions.  

We have decomposed the V light curves of all variables listed 
in Tables \ref{cfot}, \ref{abfot} and \ref{blafot} in Fourier series of 
cosines, using 6 components for the RRc variables and 6 to 15 components 
for the RRab and Blazhko stars. We list in Tables \ref{fparc}, \ref{fparab} 
and \ref{fparbla} the resulting Fourier parameters, respectively, 
where $A_{n}$ are the amplitudes of the n-th components, $A_{n1}$ are the 
amplitude ratios $A_{n}/A_{1}$, and $\phi_{n1}$ are the phase term differences 
$\phi_n - n\phi_1$.   
For the RRab and Blazhko stars we have also estimated the $Dm$ parameter, 
defined by Jurcsik \& Kov\`acs (1996) and Kov\`acs \& Kanbur (1998) as the 
maximum value among several combinations of Fourier coefficients, and 
represents a quality test on the regularity of the shape of the light 
curve. All the formulae defined by Kov\`acs and collaborators relating 
physical stellar properties to Fourier parameters are applicable and 
``reliable'' only if the {\em compatibility condition} $Dm < 3$ is met. 
In the calculation of the $Dm$ parameter all values of 
$\phi_{41} < 2$ listed in Tables \ref{fparab} and \ref{fparbla} 
have been increased by $2\pi$ in order to bring them to the 
typical value range of $\phi_{41}$. 
The $Dm$ parameter is listed in the last column of Tables \ref{fparab} 
and \ref{fparbla}. 

Since one of the aims of this study is to get an idea of how 
a Blazhko star changes its observable parameters during the Blazhko 
modulation cycle, we have applied the Fourier analysis separately also to 
the Kal98 and Car98 $V$ light curves, when available, and listed 
the corresponding Fourier parameters in Table \ref{fparbla}.  
By comparing with Kal98 results on common non-Blazhko stars, we note 
that our estimates of $Dm$ are systematically smaller, which is 
probably due to the fact that we have applied a different set 
of equations (Kov\`acs \& Kanbur 1998, their Table 2) with respect to Kal98. 
However, all values listed in Table  \ref{fparbla} have been calculated 
with the same procedure, and are therefore homogeneous and comparable. 

An inspection of the $Dm$ values derived for both RRab and Blazhko 
variables leads us to two important considerations:

\noindent i) In Table \ref{fparab}, of the 67 RRab stars that we have selected 
for our analysis, 12 have $Dm > 5$, 11 have  $3 < Dm < 5$, and the remaining 
44 have $Dm < 3$. Therefore, in spite of the severity of our initial selection 
criteria on photometric quality, only 2/3 of our sample meet the strictest 
requirement ($Dm < 3$) set by Kov\`acs \& Kanbur (1998) for the application of 
their relations to derive ``reliable'' stellar physical parameters.  
This fraction goes up to 82\% if this requirement is slightly relaxed 
($Dm < 5$). 

\noindent ii) In Table \ref{fparbla}, of the 38 Blazhko stars selected for our 
analysis, 14 stars have $Dm<3$ in at least one Blazhko phase and $Dm<5$ 
at all other phases; 2 stars have $Dm<5$ at all detected Blazhko 
phases; of the remaining stars,  17 show at least one 
value larger than 5 at some Blazhko phase, {\em as well as} values smaller 
than 5 and even 3 at some other Blazhko phase. 
Altogether, we detect as much as $\sim$ 40\% (63\%) cases of  
Blazhko stars with  $Dm<3 ~(5)$, {\em with no correlation with the 
amplitude of the corresponding light curves}.  

Recall that Szeidl (1976, 1988) found that Blazhko RR Lyrae stars, 
when observed at the Blazhko modulation phase corresponding to the largest 
light curve amplitude, behave like regular stars. In order to verify whether 
there is any ``regular'' phase during Blazhko modulation,  Jurcsik et al.\ 
(2002) have re-examined 4 field Blazhko RR Lyrae variables: RR Lyr at 6
Blazhko phases, RV UMa at 8 Blazhko phases, AR Her at 4 Blazhko phases,
and RS Boo at 3 Blazhko phases. They set rather strict criteria for 
regularity, one of them being $Dm<2$, and  conclude that 
only stars with small Blazhko amplitude modulation can show regular 
light curves at some Blazhko phases, a rather small fraction ($\le$20\%) 
of the total considered cases.  
Therefore, a criterion such as $Dm<2$ could be quite effective at detecting 
non-regular (Blazhko) stars, but is also more severe than the requirement 
$Dm<3$ set by Kov\`acs \& Kanbur (1998) for the applicability of their 
formulae, and so excessively penalizing (e.g. only 27 of our 67 supposedly 
regular RRab variables would meet this criterion). 
On the other hand, by adopting the criterion $Dm<3$, we find that 
a relatively large fraction of known Blazhko stars show regular curves at some 
Blazhko phase. We conclude  that {\em the condition $Dm<3$  is not an effective 
indicator of the Blazhko behaviour}, contrarily to the conclusions 
reached by Jurcsik \& Kov\`acs (1996). 

In the following sections we shall apply  Kov\`acs \& Kanbur (1998) relations 
to all our target stars, but consider the results only when $Dm<5$. 
We have adopted a slightly relaxed criterion with respect to the 
original one in order to improve the statistics, after we have verified that 
this does not lead to any significant difference in the resulting 
physical parameters.

\subsection{Pulsation modes}

The first use of the Fourier parameters is aimed at identifying the pulsation 
mode of the stars, which may be ambiguous in a few cases. 
In Fig. \ref{cacciari.fig13}, upper panel, we show  $A_{21}$ as a function of 
$\phi_{21}$, for all normal stars in Tables \ref{cfot} and \ref{abfot}. 
The fundamental  and first overtone  pulsation modes are clearly separated at 
$A_{21} \sim$  0.3. 
We note that V202, supposedly an RRab star with very long period and small 
amplitude, falls in the domain of the RRc pulsators. 
A few stars falling off the main RRc distribution (cf. Sect. 3.2) are marked: 
we find again the overluminous variables V70, V129 and V170, and 
V203 that can be an RRe pulsator, whereas V105 and V178 fall within the 
group of normal first overtone pulsators. 
In the lower panel we show the Blazhko stars in the large amplitude (filled
circles) and small amplitude (open circles) phase: they mostly   
fall in the typical RRab domain, except a few small-amplitude cases on the 
borderline and V41 in the small amplitude phase that falls clearly in the RRc 
domain (but its light curve is rather bad and the $Dm$ parameter is large).   
Therefore we conclude that, at least in the sample we have selected, all 
the Blazhko stars seem to pulsate in the fundamental mode.

\subsection{Physical properties derived from Fourier parameters}

Although the physical link between the Fourier parameters describing 
the shape of a light curve and the physical parameters of that particular 
pulsating star is not yet understood, nevertheless clear and well 
defined relations have been found empirically. 

The work by SC93  deals with first-overtone RR Lyrae variables  
using Fourier decomposition in cosine series of V light curves to estimate 
the mass, luminosity and temperature of these stars. 
The work by Kov\`acs and collaborators has focussed instead on the 
fundamental pulsators, with the exception 
of Kov\`acs (1998) who provides a relation to estimate the 
absolute magnitude $M_V$ of RRc stars. The relations defined for 
the RRab stars allow to derive parameters such as $M_V$, [Fe/H], 
intrinsic color and temperature, based on the Fourier decomposition 
in sine series of V light curves. Therefore, before applying these 
relations we have corrected our phase parameters $\phi_{21}$, $\phi_{31}$ 
and $\phi_{41}$ values  (listed in Tables \ref{fparab} and \ref{fparbla}), 
that were derived from cosine series, by --1.57, +3.14 and --4.71 respectively. 

Tables \ref{phyfc},  \ref{phyfab} and \ref{phyfbla} list the physical 
parameters we have derived for the RRc, RRab and Blazhko stars, respectively. 
We shall discuss them in more detail in the following sections.

\subsubsection{The Metallicity [Fe/H]} 

The connection between the period and the Fourier parameter $\phi_{31}$ of the 
V light curve of a RR Lyrae variable star was investigated already several 
years ago (Petersen 1984; Simon 1989, 1990), and it was initially attributed 
to a dependence on the star mass by Simon. 
From the analysis of RRc stars in five globular clusters, Clement et al.\ (1992) 
found the trend with metallicity of the $\phi_{31}$-period relation, which 
``appears to be the Sandage period shift in another guise''. 
Indeed, Sandage (2004a) reaches the same conclusion from the detailed analysis 
of 55 field RRab stars. 
So, although the physical significance of the $\phi_{31}$-period 
relation may still be elusive, its connection with metallicity is well 
defined. This relation can therefore be used to derive estimates of 
metallicity for variable stars once their periods and V light curves are known 
with sufficiently good accuracy.   

From the analysis of 272 V light curves of RRab stars taken in nearly equal 
number from the Galactic field, the Galactic globular clusters and the 
Sculptor dwarf galaxy, and using high-dispersion spectroscopy for the
metallicities, Jurcsik (1998) derived the relation (for RRab stars):

\noindent $[Fe/H]=-5.038-5.394P+1.345\phi_{31}$ (r.m.s. error of the fit 0.14 
dex) \hfill (10)

\noindent The zero-point of this new metallicity scale compares to the 
traditional ZW scale as $[Fe/H]_{J}=1.431[Fe/H]_{ZW}+0.88$ (Jurcsik 1995), 
and for M3 yields  $[Fe/H]_{J}$=--1.50, therefore the application of eq. (10)  
should give consistent values of metallicity with our assumptions. 

We list these [Fe/H] determinations in Tables \ref{phyfab} and \ref{phyfbla}. 
The average values of metallicity we derive, considering only the RRab (or 
Blazhko) stars 
with $Dm<5$, are $<[Fe/H]_J>$=--1.39$\pm$0.11 for the 45 normal RRab stars, 
--1.40$\pm$0.14 for the 6 longP/overluminous stars, and --1.17$\pm$0.05 for 
the 4 low amplitude/suspected Blazhko stars. \\  
Here a few comments can be made: \\
i) The zero-point of eq. (10) obviously depends on the entire calibrating 
sample, and the spectroscopic value of metallicity for M3 (from RGB stars) is 
$\sim$0.1 dex more metal poor than the value predicted by eq. (10). 
The comparison with Sandstrom et al.\ (2001) spectroscopic abundances of 
29 RR Lyraes, that was done by Jurcsik 
(2003), is hardly useful since these abundances are based on low resolution 
spectra and their accuracy is quite poor. 
Direct spectroscopic abundances of RR Lyrae stars in the LMC have been obtained 
by Gratton et al.\ (2004) and compared with [Fe/H] estimates from
eq. (10). Because of large errors and a possible spread in metallicity, and 
the presence of a significant number of outliers, the comparison of the 
spectroscopic and Fourier metallicity determinations does not appear very 
conclusive, although some general agreement is indeed evident. 
A recalibration of the [Fe/H]-period-$\phi_{31}$ relation has been performed 
by Sollima et al.\ (2004) using a non-parametric fitting routine based on local 
polynomial surface fitting on a database of 287 RRab variables in 18 globular 
clusters. The cluster metallicities are taken from KI, and the r.m.s. of the 
fit is $\sim$0.16 dex. Based on the 25 calibrating clusters used by Jurcisk
(1995), we have compared these two metallicity scales and found that they are 
roughly linearly related as $[Fe/H]_J=0.8[Fe/H]_S-0.2$. We show in 
Fig. \ref{cacciari.fig14} the histograms of the metallicity distributions 
derived from eq. (10) (shaded area) and from Sollima et al.\ (2004) 
recalibration (solid line) for the 45 regular RRab stars considered above. 
The average value from this recalibration is $<[Fe/H]_S>$=--1.43$\pm$0.07, 
in good agreement with the average value from eq. (10) and indicating that M3 
lies indeed $\sim$0.1 dex off the calibration defined by a 
large number of globular clusters, irrespective of the fitting method and 
metallicity scale. The recalibrated metallicity distribution appears to be 
narrower and somewhat different in shape than the distribution from eq. (10). 
In the case of M3 these differences are well below the errors of the 
respective determinations, and we may conclude that the two distributions are 
on average comparable, within the respective errors. However, these differences 
could be much more significant in the case of a composite population with an 
intrinsic spread in metallicity, and then it would be important to assess which 
fitting method and/or metallicity scale yields the metallicity
distribution that best reproduces the spectroscopic one. \\
ii) The r.m.s. errors we estimate from our sample of 45 best stars  are well 
below  the intrinsic accuracy of the fits given by Jurcsik (1998) and Sollima 
et al.\ (2004).  This argues against the recent claim by Jurcsik (2003) of the 
existence of a metallicity dispersion among the variable stars of M3. 
Similar conclusions 
are reached by Sandstrom et al.\ (2001), who find that ``the compositions of 
RR Lyrae stars in M3 are uniform within [their] sample and consistent with the 
compositions of M3's giants'', and by KI who find a dispersion of 
$\sigma$=0.03 dex for the FeII abundances averaged over 23 giant stars.  \\
iii) A further check of the above results using the metallicity data listed 
in Table \ref{phyfbla} for the Blazhko stars shows that the average of all 
values, irrespective of the Blazhko phase, is --1.37$\pm$0.30, and 
the average values corresponding to the small-amplitude and 
large-amplitude phases are --1.29$\pm$0.37 and --1.42$\pm$0.24, respectively.  
Note that the r.m.s. errors are significantly larger than for regular 
variables. Although the differences in $<[Fe/H]>$ may not be statistically 
significant, given the large errors, they nevertheless indicate that 
Blazhko stars at large-amplitude Blazhko phase are quite similar 
to regular pulsators, whereas at small-amplitude Blazhko 
phase the light curves are more likely distorted (in spite of $Dm\le5$), and 
tend to overestimate the metallicity. Therefore, {\em including in the sample 
unrecognised Blazhko stars can produce distorted metallicity distributions}. 
An inspection of Jurcsik (2003) target list reveals that 10 out of her 29 
stars are Blazhko, one is a suspected Blazhko and one has $Dm>5$ in our data. 
If we exclude these stars from the average, we find $<[Fe/H]>$=--1.42$\pm$0.09 
using our estimates of metallicity, and --1.37$\pm$0.13 using  Jurcsik's 
estimates. 
This result does not show any evidence of a metallicity spread, and we think 
that the inclusion of Blazhko stars in the sample is what led Jurcsik (2003) 
to an incorrect conclusion. \\       
iv) Sandage (2004a) estimated the effect of evolution off the ZAHB on the 
metallicity determination.  His conclusion is that evolution produces a noise 
in [Fe/H], in the sense that a $\Delta\log\,P$=0.10 dex at fixed $\phi_{31}$ 
due to evolution at constant [Fe/H] would generate an error 
$\Delta$[Fe/H]=0.67 dex. 
Since, as we have seen in Sect. 3.2, the period shift at fixed amplitude  
between the main body of the RRab stars and those labelled 
``longP/overluminous'' is $\Delta\log\,P \sim$ 0.06,  we should expect a 
systematic overabundance for these evolved stars of $\sim$0.40 dex: 
however, we find none. 
In his analysis, Sandage could see and identify correctly the 
monotonic variation within the instability strip of period, amplitude and 
$\phi_{31}$, as we also see in our Fig. \ref{cacciari.fig5}. 
On the assumption of a unique amplitude-$\phi_{31}$ relation, he  
deduced that the same period shift that occurs in the period-amplitude plane 
would occur also in the period-$\phi_{31}$ plane, whether due to evolution 
or to an abundance difference. The problem seems to lie in this assumption.  
As one can see in Fig. 2 of Jurcsik et al.\ (2003), and more clearly  
in our Fig. \ref{cacciari.fig15}, at fixed [Fe/H] the period 
shift due to evolution that stands out in the period-amplitude plane 
disappears in the period-$\phi_{31}$ plane, where all stars follow the same 
relation irrespective of their evolutionary status. 
This applies also to the RRc stars, that follow their own specific relation, 
different from the RRab's.  
This effect could not be seen in Sandage's target sample because of the lack 
of clearly identifiable evolved stars.  
Therefore, if the period-shift (at fixed amplitude) that appears in the 
period-amplitude plane is due to metallicity differences, the same period-shift 
(at fixed $\phi_{31}$) will appear also in the period-$\phi_{31}$ plane,  
and the variation in period will produce a variation in metallicity as 
estimated by Sandage. 
If, however, the period-shift in the period-amplitude plane is due to 
evolution, there is no corresponding shift in the period-$\phi_{31}$ 
plane.
So {\em evolution off the ZAHB will produce no noise on metallicity 
determinations}.

\subsubsection{The Intrinsic Colors (B--V) and (V--K)} 

The possibility of estimating intrinsic colors from the Fourier parameters 
of the RRab stars has some important implications, allowing e.g. to estimate 
reddening and temperature of these stars. These color indices, as defined by 
Jurcsik (1998), are the  differences of the {\em magnitude-averaged} absolute 
brightnesses.  The following relations are taken from Kov\`acs \& Walker 
(2001): 

\noindent $ (B-V)_0 = 0.189logP - 0.313A_1 + 0.293A_3 + 0.460 $ \hfill (11)

\noindent $(V-K)_0=1.257P-0.273A_1-0.234\phi_{31}+0.062\phi_{41}+1.585$
\hfill (12)

\noindent 
These intrinsic colors, that are listed in col. 5 and 6 in Tables \ref{phyabc} 
and \ref{phybla}, 
can be used to estimate reddening and temperatures. However,  before any 
practical application it is worth checking how they compare with their 
observed counterparts, e.g. (B--V)$_{mag}$ from CC01 that have been 
reported for convenience in column 9 of Table \ref{abfot}, and (B-V)$_S$  
that have been used to derive the temperatures in Sect. 5. 
We show in Fig. \ref{cacciari.fig16} the histograms of these three color distributions 
for the RRab stars, and note that the (B--V)$_0$  distribution is somewhat 
compressed and slightly blue-shifted with respect to the (B--V)$_{mag}$ 
distribution. Whereas the blue-shift corresponds, correctly, to a reddening 
of about 0.01 mag, the reduced width is a distortion that becomes
even more evident when compared to the (B--V)$_S$ color distribution. 
In particular, the red and blue edges of the (B--V)$_0$  distribution occur 
at about 0.38 and 0.28, respectively, instead of 0.41 and 0.24.
This distortion in the shape of the (B--V)$_0$ color 
distribution will have some consequence on the temperature determination, as 
we shall see below. 
 
\noindent {\bf Reddening} \\
From the difference of the observed (B--V)$_{mag}$ colors  and the intrinsic 
$(B-V)_0$ colors derived from eq. (11) we derive 
a mean reddening $E(B-V)=0.007\pm0.013$ from the RRab stars with $Dm<5$. 

If we consider the 19 RRab stars with infrared data from L90 and $Dm<5$, and 
compare the observed (V--K) colors with those derived from eq. (12), 
we obtain an average $E(V-K)=-0.011\pm0.063$, hence E(B--V)=--0.004. 
This result has a larger r.m.s. error than that from $(B-V)_0$ colors, 
possibly due to the smaller number of stars used for this estimate.    

Both results are compatible, within the respective errors, with the 
reddening estimated in Sect. 4.1.  
 
\noindent {\bf Temperatures} \\
The expressions used above to derive the intrinsic colors of the RRab 
variables can be used to calculate their effective temperatures: 

\noindent $\log\,T_{eff}(B-V)=3.930-0.322(B-V)_0+0.007[Fe/H]$ \hfill (13)

\noindent $\log\,T_{eff}(V-K)=3.929-0.1112(V-K)_0-0.0032[Fe/H]$ \hfill (14)

\noindent Eq. (13) is taken from Kov\`acs \& Walker (2001, eq. 11), where the 
gravity is assumed as $\log\,g=2.75$.  A variation $\Delta\log\,g=\pm0.1$ dex 
would reflect on the temperature as $\Delta T_{eff}=\pm$28 K. 
This color-temperature relation was calibrated on Castelli et al.\ (1997) models 
that are very similar to the C99 models discussed in Sect. 4.2, and is 
based on a different definition of mean color for RR Lyrae variables. 
Eq. (14) is taken from Jurcsik (1998). These temperatures are listed in 
Table \ref{phyfab}, col.s 7 and 8 respectively.   
For both eq. (13) and (14) the value for [Fe/H] is taken from Eq. (10), 
for self consistency within the Fourier analysis. 
This value is {\em on average} $\sim$ 0.1 dex higher than the value adopted 
for M3 from high dispersion spectra. The use of the fixed adopted value 
--1.5 for each star would make little difference on the average temperatures, 
i.e.  about --10 and +5 K from (B--V) and (V--K), respectively, with no 
systematic effects across the instability strip. 

For the RRc-type variables, the values of temperature listed in Table
\ref{phyfc} have been calculated from the relation (SC93):

\noindent $logT_{eff} = 3.7746 - 0.1452logP + 0.0056\phi_{31}$ \hfill (15)

\noindent where the dependence on color is replaced  by the dependence on
period. These temperatures are not on the same absolute scale as those derived 
from eq. (13) and (14). 

The comparison of these temperatures with those obtained using  the SF 
temperature scale and the (B--V)$_S$ colors is shown in Fig. \ref{cacciari.fig17}. 
We can see that for the RRc variables the Fourier temperatures are about the 
same as the SF temperatures at the hot end of the distribution, and become 
progressively hotter till about +500 K at the cool end. 
For the RRab stars the difference $\Delta T_{eff}$(Fourier--SF) varies nearly 
linearly from about --200 K at the hot end of the distribution to about +50 K 
at the cool end (but the coolest stars could not be used for this test because 
they have $Dm>5$ and their Fourier temperatures are not reliable). 
This effect is due to the reduced red tail of the (B--V)$_0$  distribution. 
For the RRc stars the only parameter besides period in eq. (15) is $\phi_{31}$,  
so the shape of the $\phi_{31}$ distribution must be the source of this 
effect. 
As a consequence of these differences between the Fourier and SF temperatures, 
the Fourier temperature ranges are compressed, in particular they are reduced 
to about 60\% (RRab) and 40\% (RRc) of the corresponding SF temperature ranges 
from (B--V)$_S$  colors.

\begin{figure}
\plotone{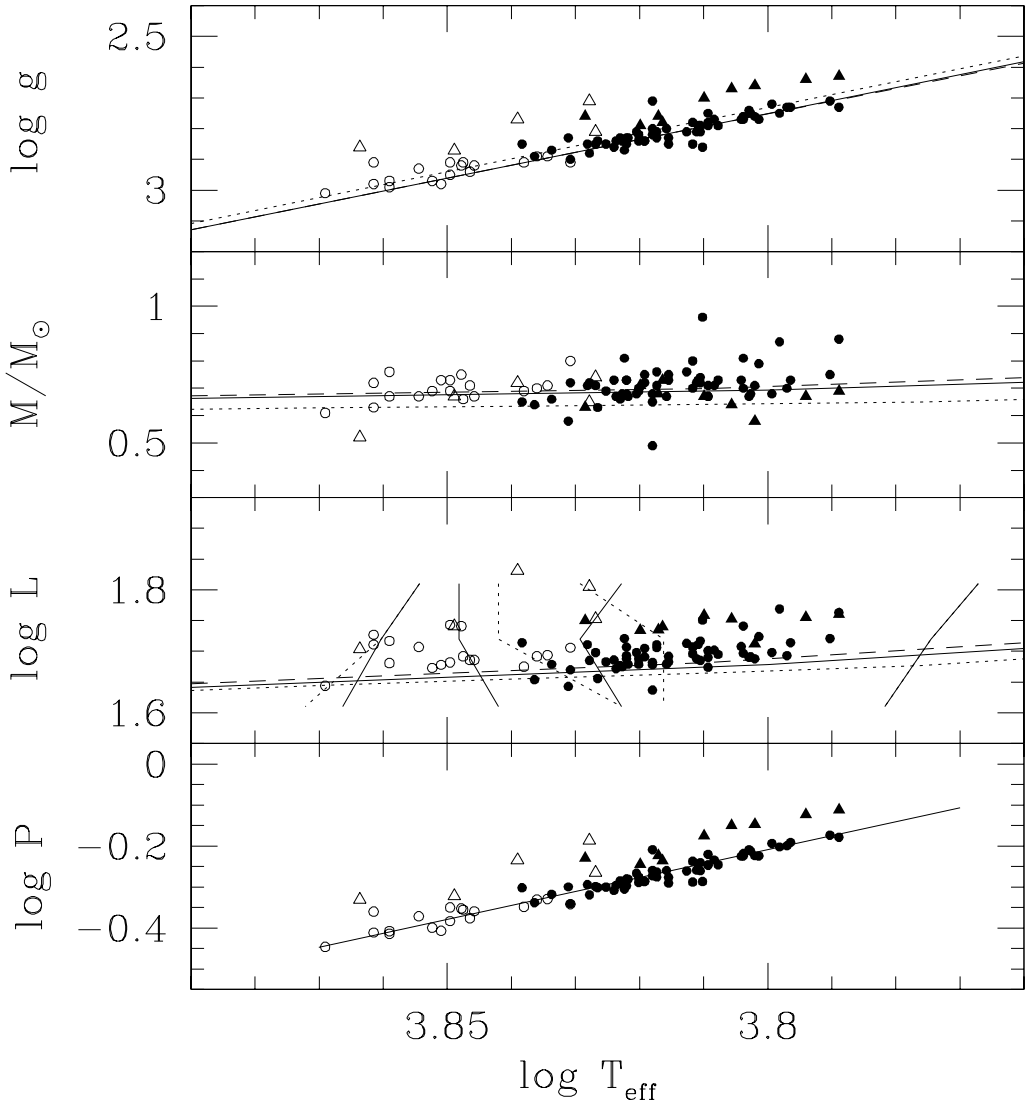}
\caption{Period, luminosity, gravity and mass are shown as a function of 
temperature for all our regular RRc and RRab stars (cf. Sect. 4),  
indicated as open and filled circles, respectively.
Evolved stars are shown as open and filled triangles.  
The line in the bottom panel is the same as in Fig. \ref{cacciari.fig8}. The 
lines in the three upper panels indicate theoretical ZAHB models, for 
comparison: solid line (Sweigart 1997, [Fe/H]=--1.53, no helium mix), 
dotted line (VandenBerg et al.\ 2000,  [Fe/H]=--1.54), and dashed line 
(Straniero et al.\ 1997, [Fe/H]=--1.63).  The nearly vertical lines in the 
$\log\,L-\log\,T_{eff}$ plane show the theoretical limits of the 
instability strip (from Bono et al.\ 1995) for mass 0.65 M$_{\odot}$ (solid 
lines) and 0.75 M$_{\odot}$ (dotted lines), i.e. from left to right: first 
overtone blue edge, fundamental blue edge, first overtone red edge, and 
fundamental red edge (where the solid and dotted lines coincide) . 
\label{cacciari.fig9}}
\end{figure}

\begin{figure}
\plotone{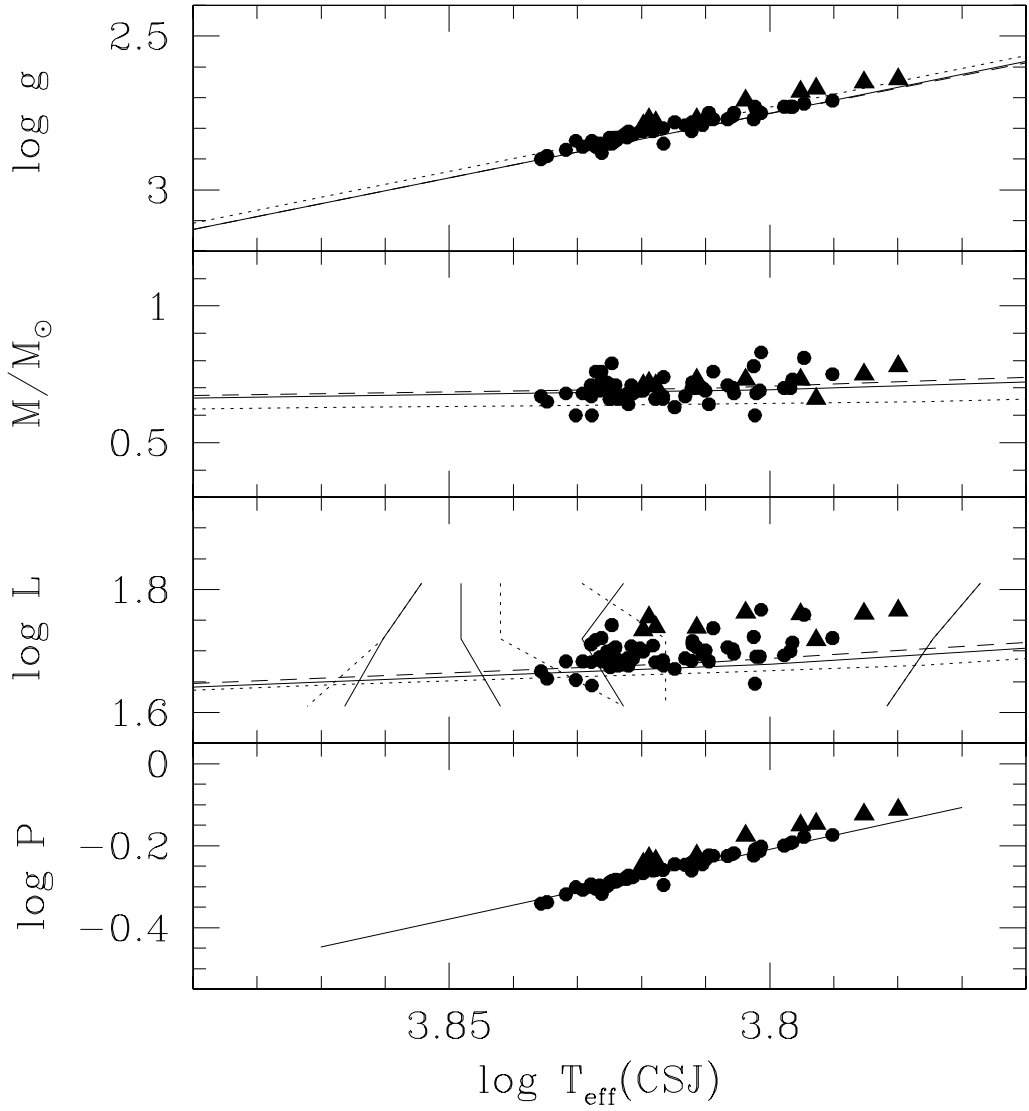}
\caption{Same as Fig. \ref{cacciari.fig9} with the temperatures derived from the 
CSJ temperature calibration (RRab stars only). 
\label{cacciari.fig10}}
\end{figure}

\begin{figure}
\plotone{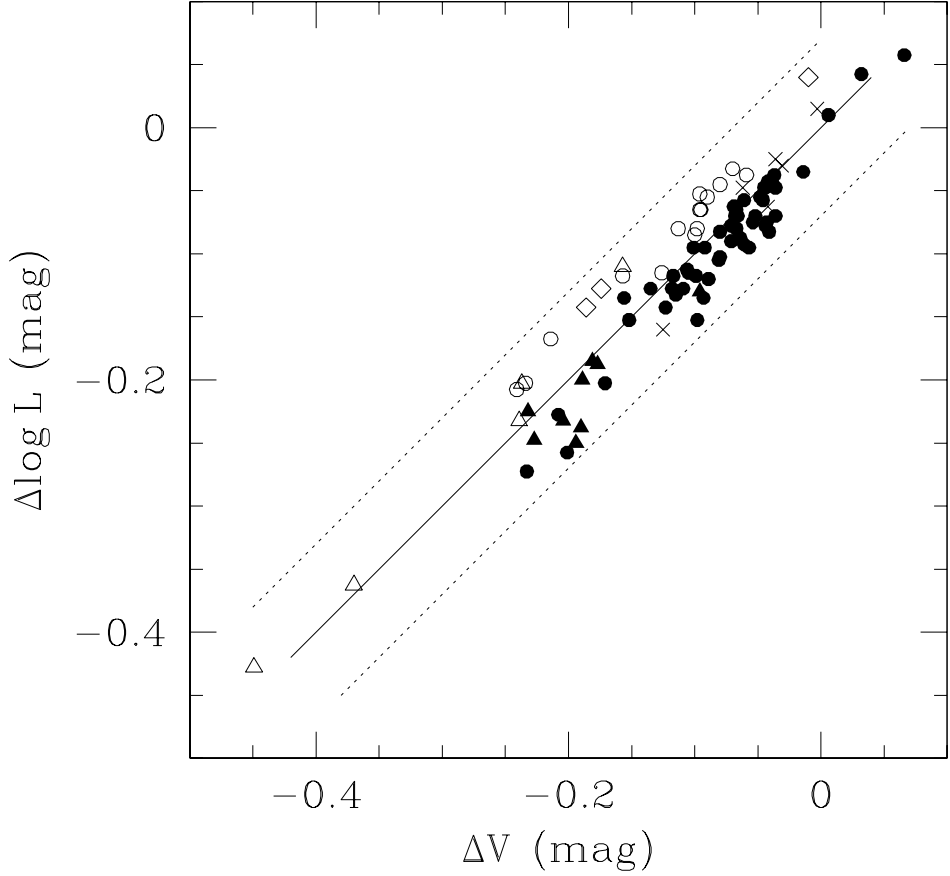}
\caption{$\Delta \log\,L$ vs. $\Delta V$ for all the RRc and RRab stars listed 
in Tables \ref{cfot} and \ref{abfot}. $\Delta \log\,L$ and $\Delta V$ are the 
calculated and observed offsets with respect to the reference ZAHB level that 
we observe at V=15.72 and calculate at $\log\,L\sim$1.66 at mid range 
color/temperature (i.e.\ B--V=0.27/$\log\,T_{eff}$=3.83). 
As in previous figures, open and filled symbols represent 
RRc and RRab stars respectively, triangles indicate overluminous (evolved) 
stars, diamonds indicate short period small amplitude RRc stars, and crosses 
indicate low-amplitude suspected Blazhko stars. The solid line represents 
the relation of slope 1, and the two dotted lines are shifted by $\pm$0.07 mag. 
\label{cacciari.fig11}}
\end{figure}

\begin{figure}
\plotone{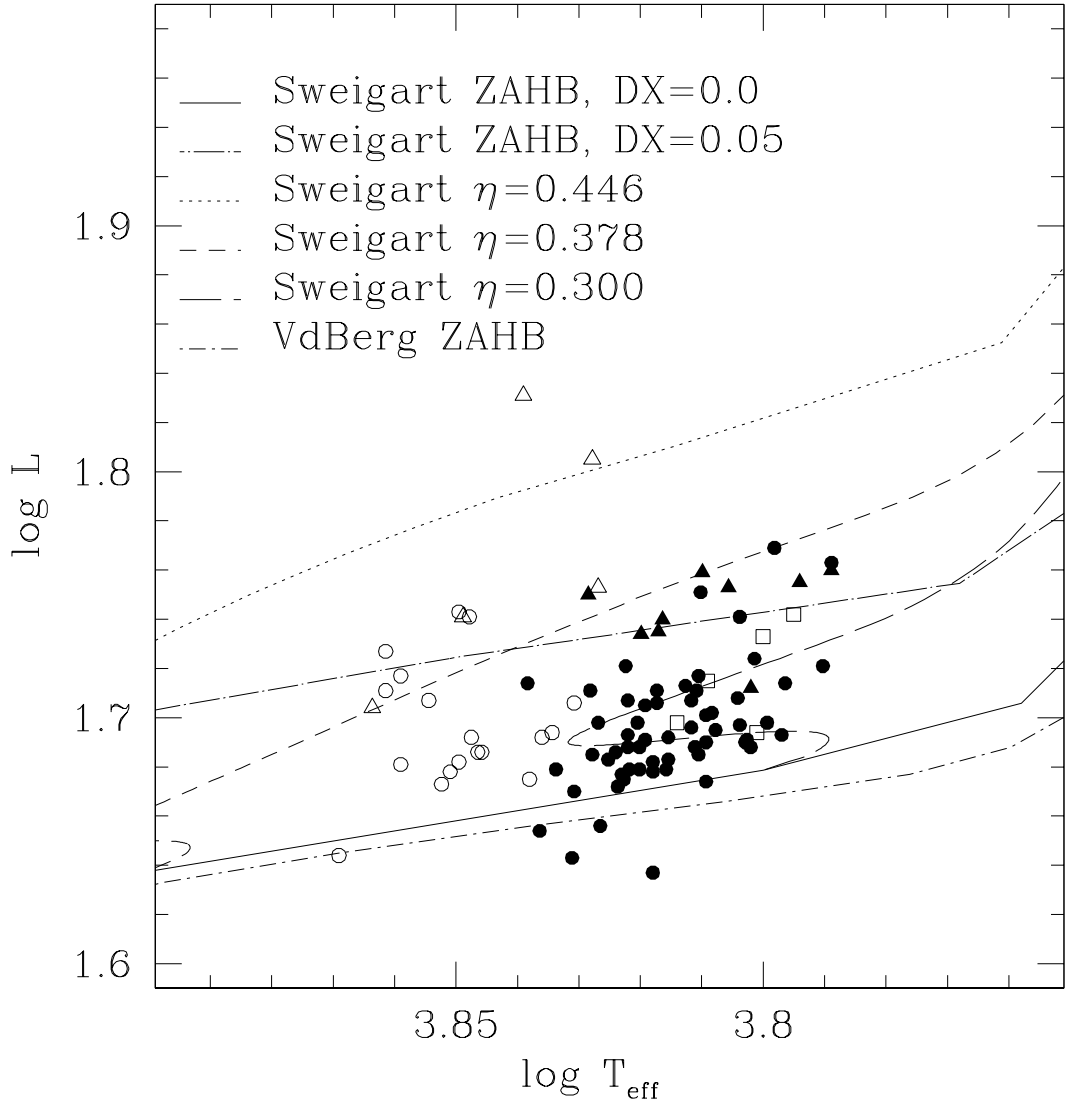}
\caption{HR diagram of the RRc and RRab stars, and comparison with various 
sets of evolutionary models.
First overtone and fundamental pulsators are shown as empty and filled circles,
respectively. Overluminous/evolved stars are shown as empty and filled 
triangles. For completeness, the 5 evolved Blazhko stars at the phase of their 
maximum light curve amplitude are also shown (empty squares).  
\label{cacciari.fig12}}
\end{figure}

\begin{figure}
\plotone{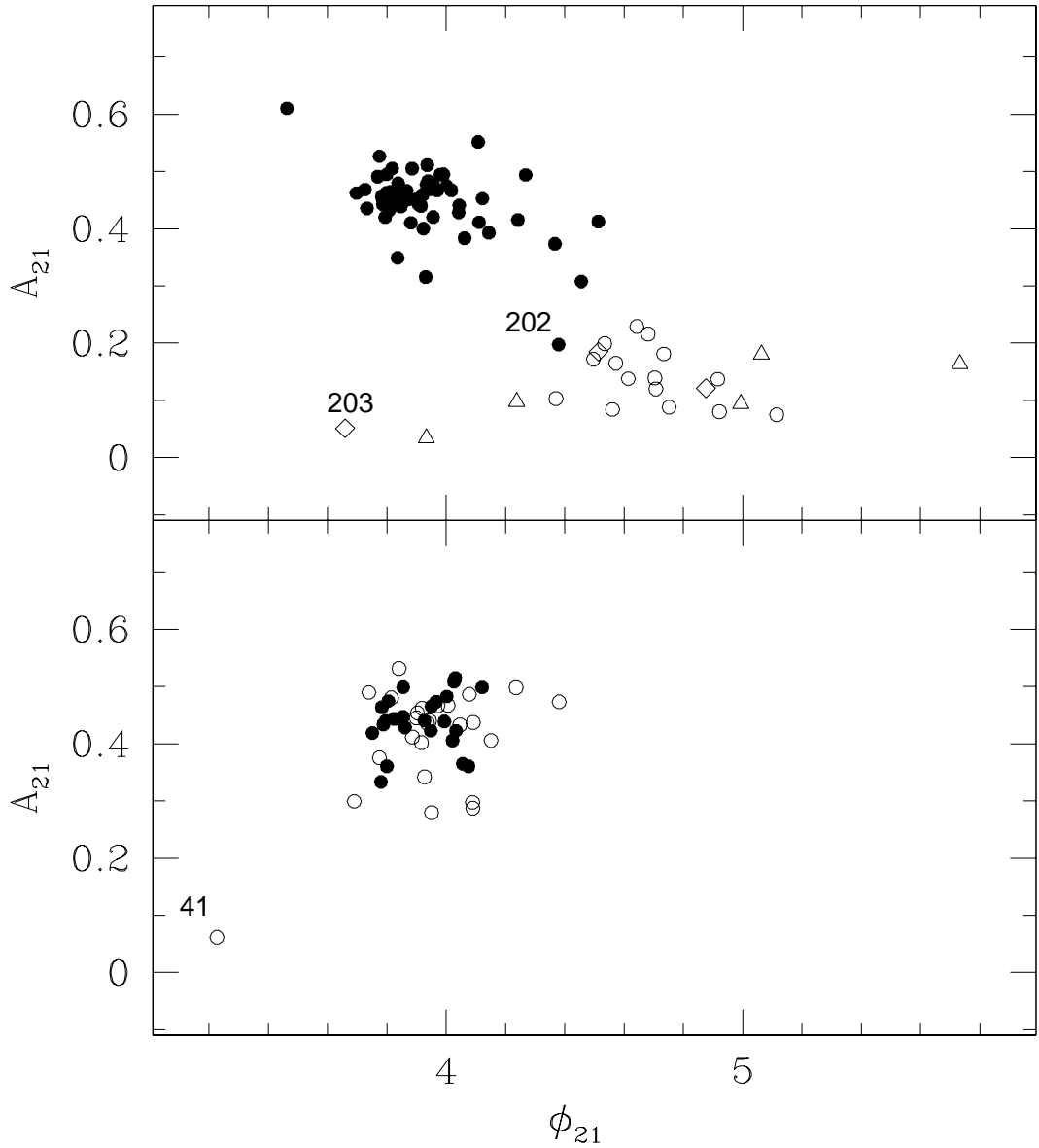}
\caption{Upper panel: Fourier parameters $A_{21}$ as a function of $\phi_{21}$ 
for our sample of  RRab (filled circles) and RRc (open circles) stars. 
The dividing line between fundamental and first overtone pulsators occurs at 
$A_{21}\sim$0.3, and V202 (the longest period RRab variable of our sample) 
falls in the RRc area. 
Of the three RRc stars V105, V178 and V203 (shown as diamonds), that are 
suggested as candidate second overtone pulsators, only V203 deviates 
significantly from the main RRc distribution. The three RRc stars shown as 
open triangles are V70, V129 and V170, with unusually long periods and bright 
magnitudes: they fall in the area of the first overtone pulsators, but off 
the main RRc distribution.  
Lower panel: same as above, for Blazhko stars in the large amplitude 
(filled circles) and small amplitude (open circles) Blazhko phase (lower 
panel). 
\label{cacciari.fig13}}
\end{figure}

\begin{figure}
\plotone{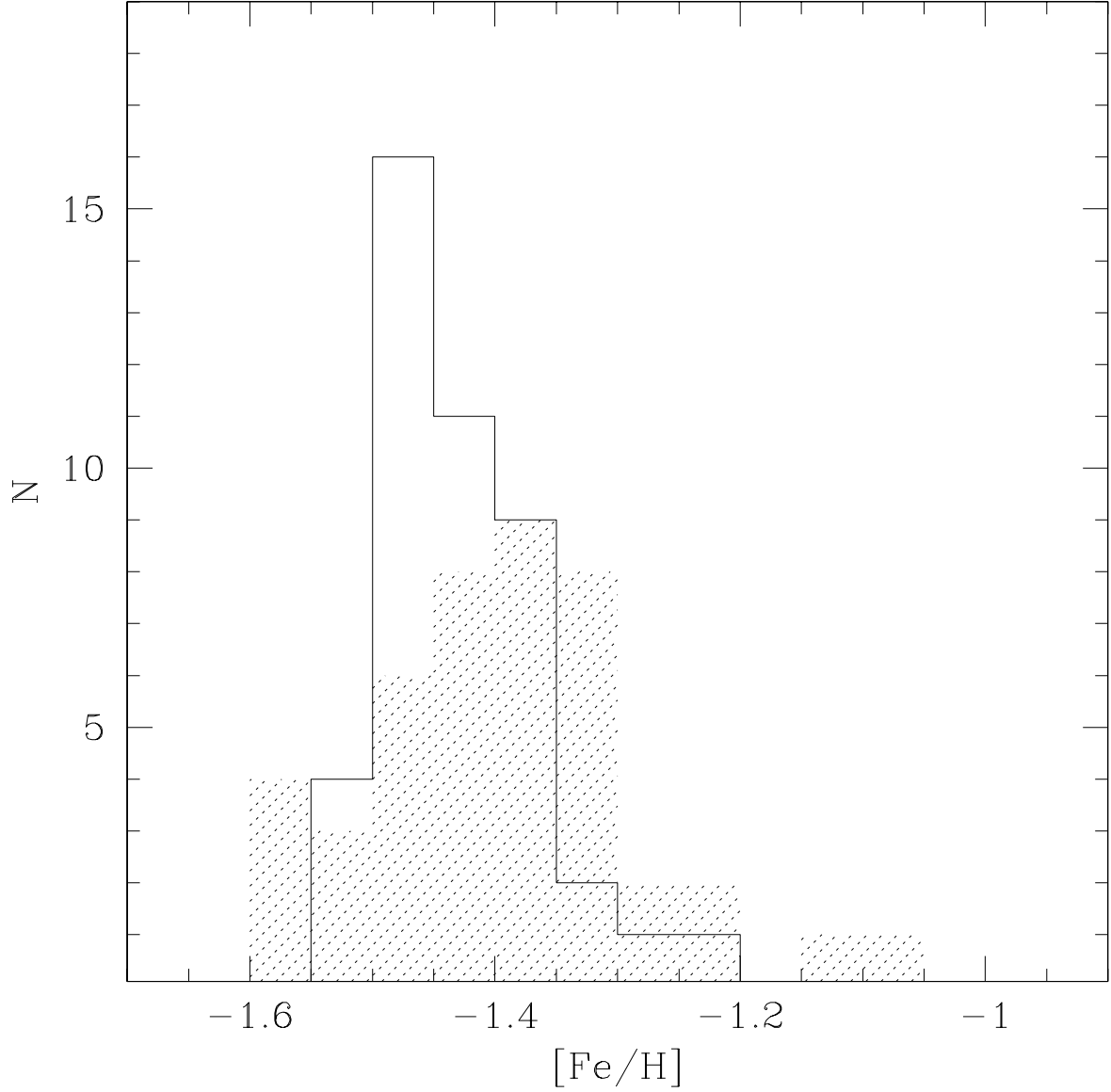}
\caption{Metallicity distributions of the 45 regular RRab stars with $Dm<5$, 
derived from eq. (10) (shaded area) and from Sollima et al.\ (2004) 
recalibration of the [Fe/H]-period-$\phi_{31}$ relation (solid line). 
The two distributions yield similar average [Fe/H] values (--1.39$\pm$0.11
and --1.43$\pm$0.07, respectively), but have different (nearly specular) 
shapes (cf Sect. 6.2.1).
\label{cacciari.fig14}}
\end{figure}

\begin{figure}
\plotone{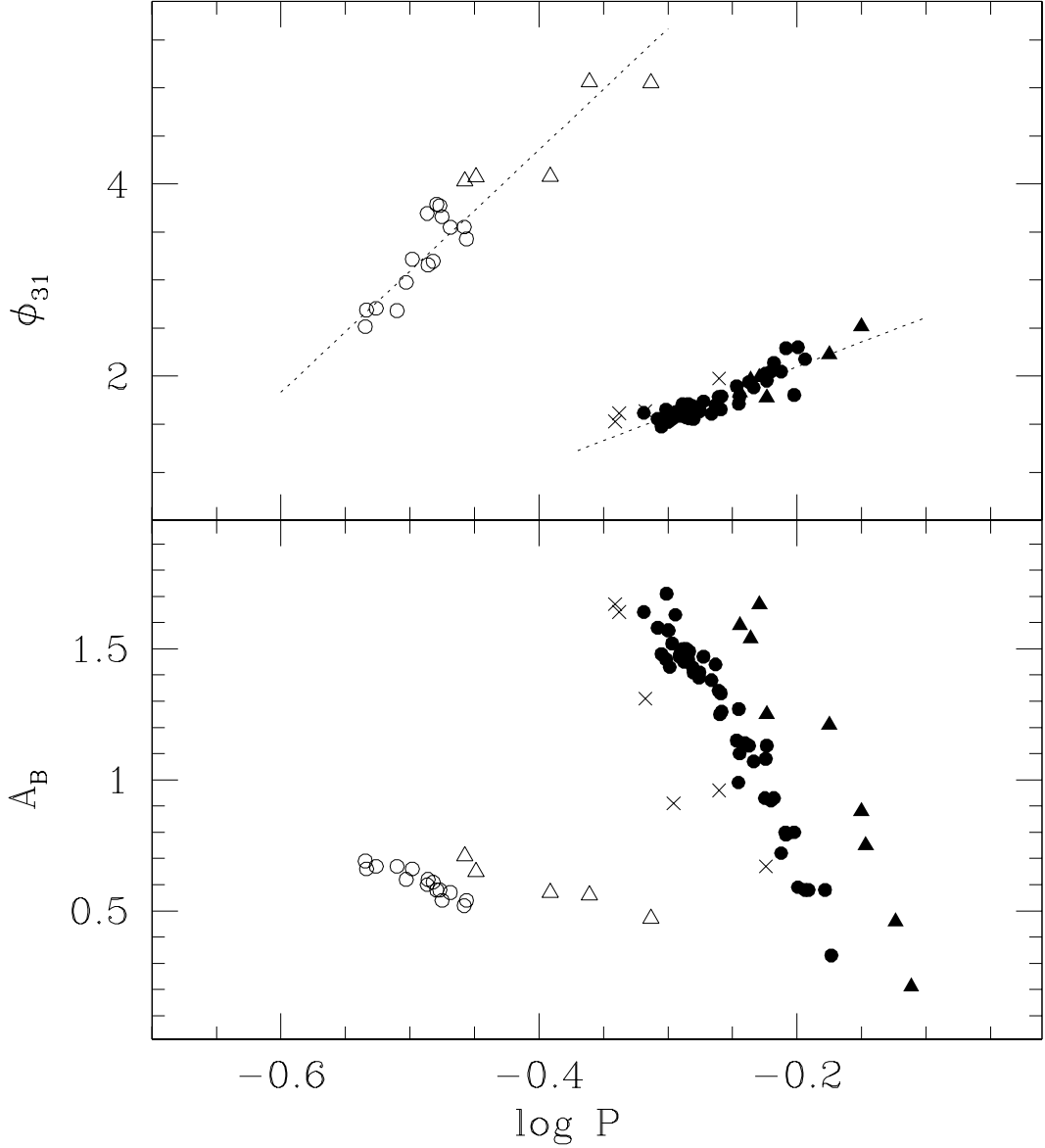}
\caption{Lower panel: blue light curve amplitude $A_B$ vs. $\log\,P$ for the 
RRc (open symbols) and RRab (filled symbols and crosses), as in Fig. 
\ref{cacciari.fig2}. Here the Blazhko stars are not shown, for the sake of clarity. 
Upper panel: the $\phi_{31}$ parameter from the Fourier series representation 
of the V light curves ($Dm\le5$ only for the RRab stars). 
The two dotted lines show the linear relations 
representing in first approximation $\phi_{31}$ vs. $\log\,P$ for the 
RRc stars ($\phi_{31}=9.403+12.619\log\,P$) and the RRab stars 
($\phi_{31}=3.124+5.128\log\,P$).   
\label{cacciari.fig15}}
\end{figure}

\begin{figure}
\plotone{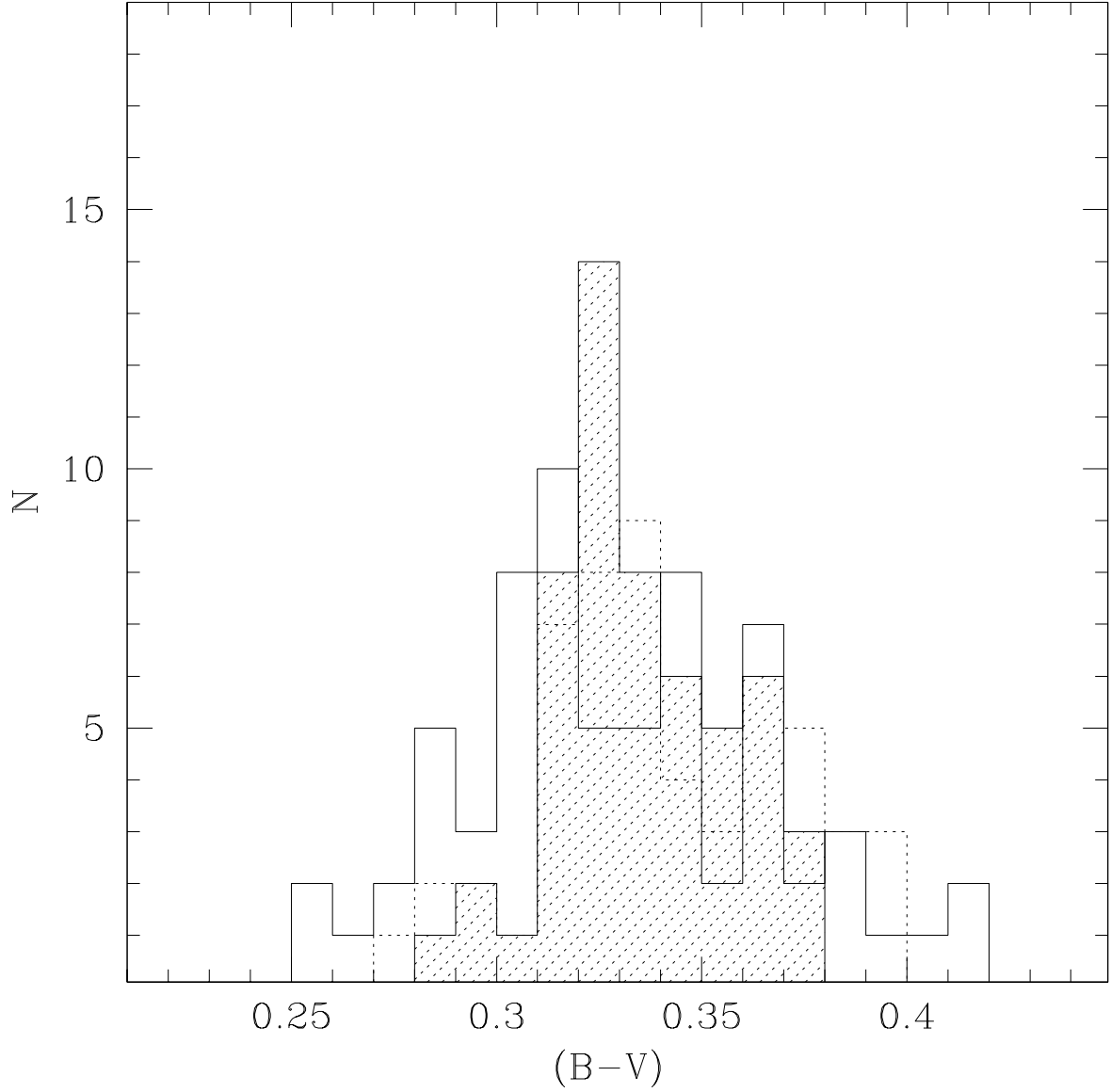}
\caption{Histograms of the observed (B--V)$_{mag}$ (dotted line) and 
adopted (B--V)$_S$ (solid line) colors from column 9 and 10 in 
Table \ref{abfot}, and intrinsic (B--V)$_{0}$ colors estimated 
from eq. (11) and listed in column 5 in Table \ref{phyfab} (shaded area). 
\label{cacciari.fig16}}
\end{figure}

\begin{figure}
\plotone{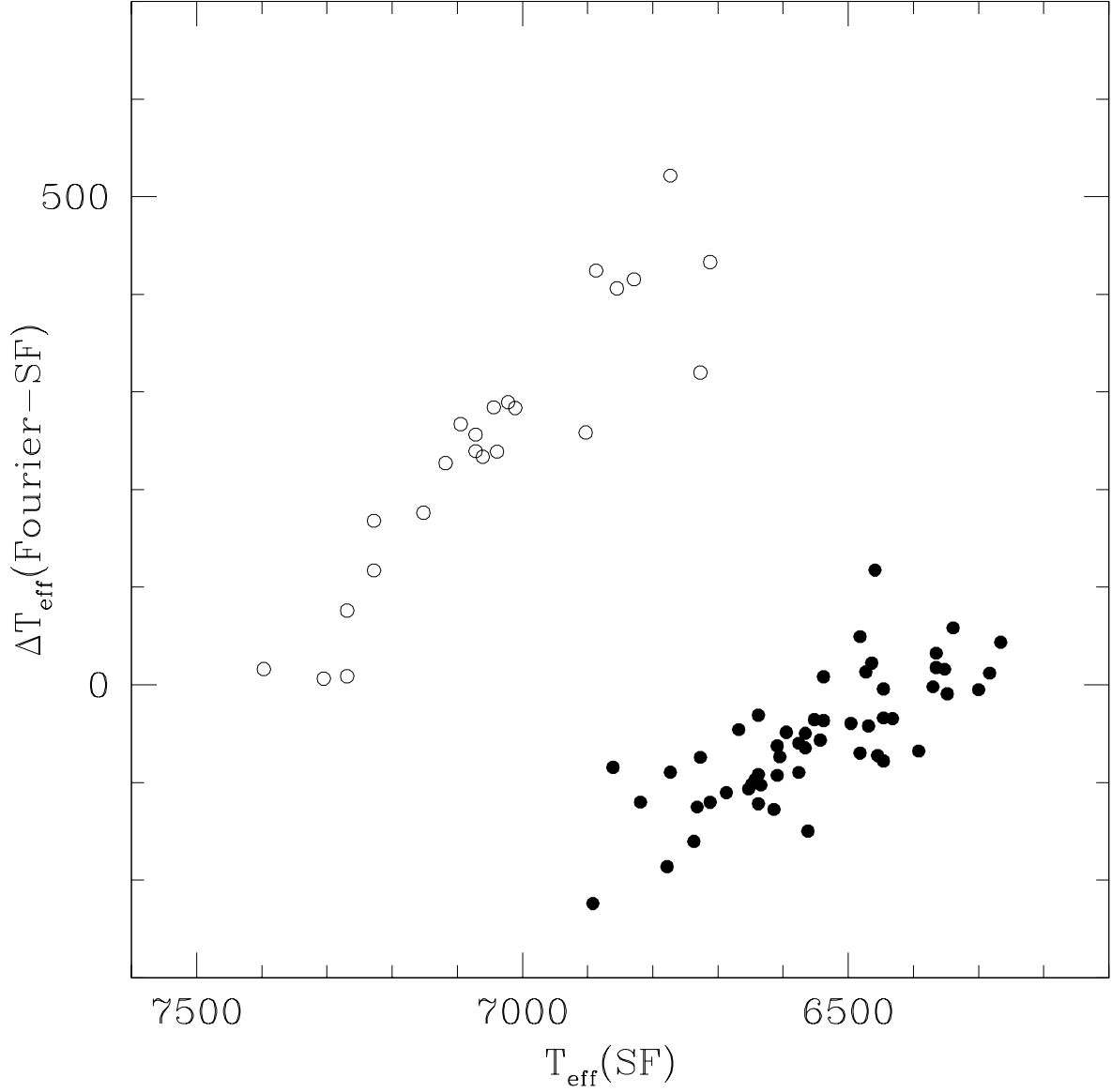}
\caption{Comparison of temperatures derived from the Fourier coefficients 
using eq. (13) for RRab stars (filled circles) and eq. (15) for RRc stars 
(open circles),  and from the (B--V)$_S$ colors and the SF color-temperature 
calibration. The Y axis shows the differences $\Delta T_{eff}$(Fourier--SF).  
\label{cacciari.fig17}}
\end{figure}

\begin{figure}
\plotone{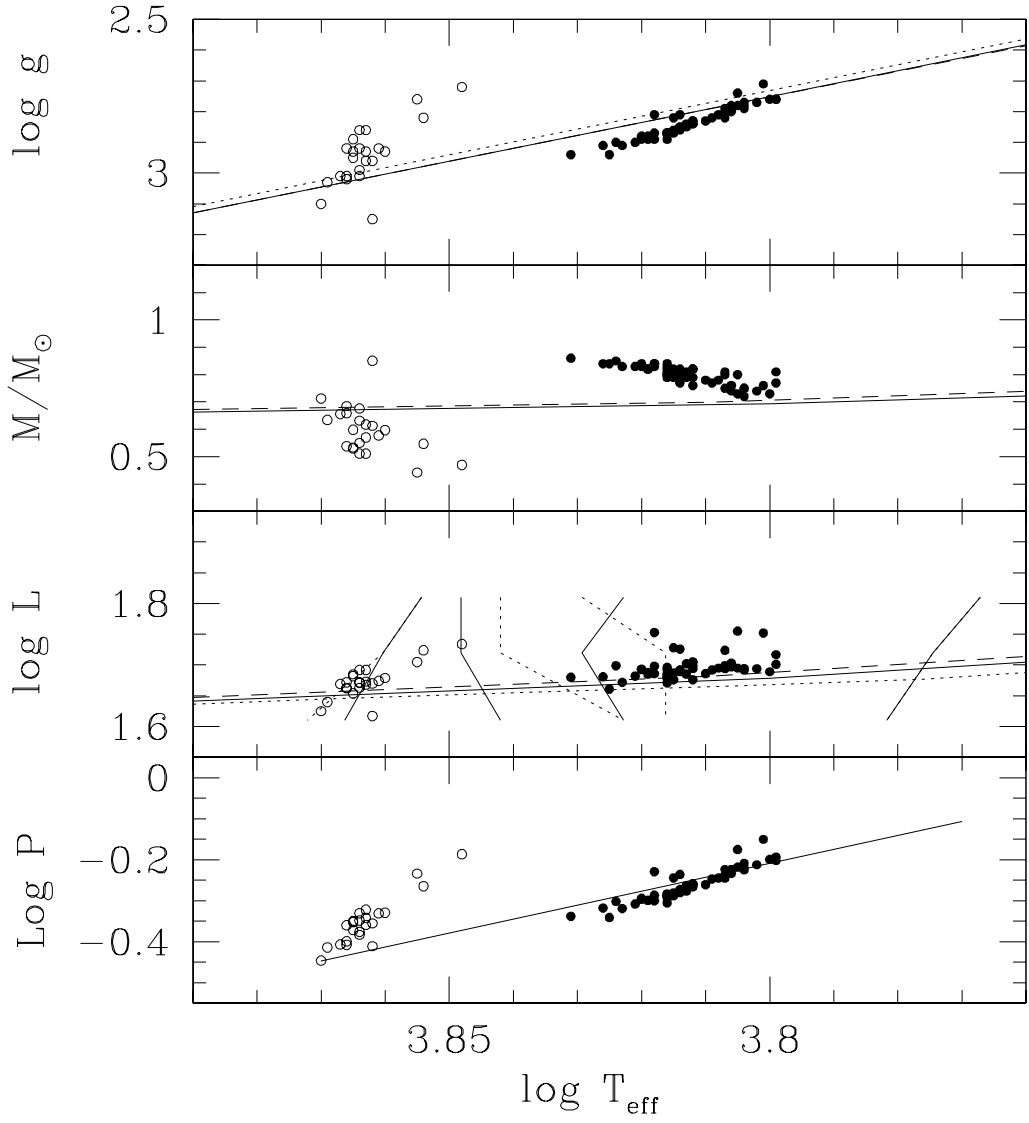}
\caption{Same as Fig. \ref{cacciari.fig9}, with the physical parameters derived from 
the coefficients of the Fourier series decomposition of the V light curves 
(cf. Sect. 6).
\label{cacciari.fig18}}
\end{figure}

We show in Fig. \ref{cacciari.fig18} (lower panel) the periods vs Fourier 
temperatures, for comparison with the periods vs SF temperatures shown in Fig. 
\ref{cacciari.fig9}. The period-temperature distributions in Fig. \ref{cacciari.fig18} do not 
follow the basic relation derived from the stellar pulsation theory,
represented by the line of slope --3.41 (cf. eq. 7), as the stars in 
Fig.  \ref{cacciari.fig9} do. Also, there is a large gap between the Fourier 
temperature distributions of the RRc and RRab stars, in clear disagreement 
with the overlap in color shown e.g. in Fig. \ref{cacciari.fig1}.

It is quite clear that the intrinsic colors derived from eq. (11) and (12) 
have a different distribution than the observed colors, and this in turn 
leads to rather questionable values of temperature via equations (13)-(15). 
{\em This method to estimate temperatures needs to be carefully re-investigated 
before the results can be used with any degree of confidence}.

\subsubsection{The Absolute Magnitude (M$_V$) or Luminosity ($\log\,L$)}

\noindent $\bullet$ {\bf RRc variables} \\
Two methods are presenty available for deriving the luminosities of RRc stars 
from Fourier parameters. One is the theoretical relation derived by SC93, 
based on hydrodynamic pulsation models matched with observations 
of globular cluster RRc stars:

\noindent $\log\,L(RRc) = 1.04\log\,P - 0.058\phi_{31} + 2.41$ (r.m.s. error 
of the fit 0.025) \hfill (16) 

\noindent where $L$ is the luminosity in solar units. 
Excluding the 4 evolved/overluminous RRc variables commented on in Sect. 3.2,  
the average luminosity of the remaining 19 RRc stars turns out to be 
$\log\,<L>=1.710\pm0.015$, that translates into $<M_V>$ = 0.44$\pm$0.03 mag 
assuming $M_{bol}$(sun) = 4.75 mag and $BC_V$=0.03 (cf. Tab. \ref{compar}). 
Kal98 found an identical result by 
applying this same relation to 5 RRc stars. However,   
the application of this method to RRc stars in 7 globular clusters led 
Clement (1996) to derive a luminosity vs. metallicity relation 
$M_V=0.19[Fe/H]+0.82$,  that would yield $M_V$=0.54 for [Fe/H]=--1.5. 
There seems to be a problem  with the definition of the zero-point with 
this formulation, but this is just a matter of calibration. 
Possibly more important, the 1$\sigma$ error associated to $<M_V>$, 
$\pm$0.03 mag, is half the value associated to the corresponding observed 
$<V>$, $\pm$0.06 mag (cf. Sect. 3.1). 

Alternatively, we can use the empirical relation by Kov\`acs (1998) 
for the intensity averaged absolute magnitude 

\noindent $M_V(RRc)=1.061-0.961P-0.044\phi_{21}-4.447A_4$ (r.m.s. error of 
the fit 0.042) \hfill (17)

\noindent where we adopt a brighter zero-point by 0.2$\pm$0.02 mag than 
the original value by Kov\`acs (1998), in order to be consistent with the 
assumptions made in Sect. 4.3.2. 
The average value we obtain is $<M_V>$=0.57$\pm$0.04 mag, and again the 
1$\sigma$ error is significantly smaller than the error on $<V>$.
In the following discussions we adopt eq. (17) to derive the Fourier luminosity 
of the RRc stars, using the BC$_V$ values listed in Table \ref{phyabc}.      


\noindent $\bullet$ {\bf RRab variables} \\
The most recent version of the relation between the intensity 
averaged $M_{V}(RR)$ and the Fourier parameters of the V light 
curve decomposition is from Kov\`acs (2002):

\noindent $M_V(RRab) = -1.876\log\,P - 1.158A_1 + 0.821A_3 + 0.43$ \hfill (18) 

\noindent The value for the constant, 0.43, was derived 
by Kinman (2002) from the Fourier decomposition of the V light 
curve of RR Lyr ([Fe/H] $\sim$ --1.4) and its absolute magnitude 
$M_V=0.61\pm0.10$ based on HST-FGS parallax (Benedict et al.\ 2002).     
The zero-point of this luminosity scale (i.e. $M_V=0.59$ at 
[Fe/H] = --1.5)  is about 0.05 mag fainter than the working 
assumptions we made in Sect. 4.3.2. 
In order to be consistent with these assumptions, we apply eq. (18) with a 
brighter zero-point by 0.05 mag (i.e. 0.38 instead of 0.43), and we obtain 
$<M_V>$=0.57$\pm$0.02 mag for the 45 regular RRab stars with $Dm<5$ in our 
sample, 0.45$\pm$0.04 mag for the 6 long period/overluminous stars, and 
0.59$\pm$0.02 mag for the 4 low amplitude/suspected Blazhko
stars. All these values are in the same luminosity scale that led to 
$<M_V>$=0.57$\pm$0.04 mag for the RRc stars above.

\begin{figure}
\plotone{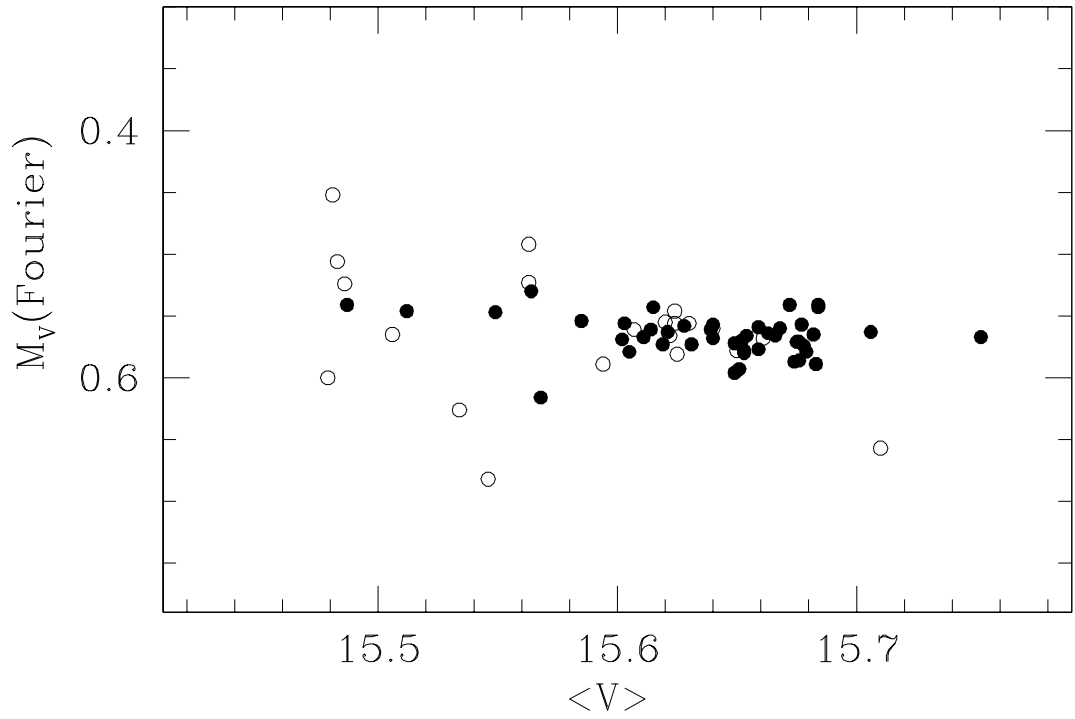}
\caption{$M_V$ vs. $<V>$, for the RRc stars (open circles) and the regular 
RRab stars with $Dm<5$ (filled circles). The values of $M_V$ have been derived 
from the Fourier parameters and eq.s (17) and (18), respectively.
\label{cacciari.fig19}}
\end{figure}

A few considerations can be made: \\
i) The dispersion of the $M_V$(Fourier) estimates 
is significantly smaller than the dispersion in the observed V magnitudes.  
We show in Fig. \ref{cacciari.fig19} a plot of the $M_V$(Fourier) values derived 
above for the RRc and RRab stars vs. the corresponding V magnitudes: 
the correlation between these two quantities is definitely flatter than one 
would expect. In particular for the RRab stars, the width of the $<V>$ 
distribution ($\sim$0.25 mag) is much larger than the width of the 
$M_V$(Fourier) distribution ($\le$0.1 mag), as also shown by the r.m.s. errors 
associated to the $<V>$  and $<M_V>$ determinations, i.e. $\pm$0.05 and 
$\pm$0.02 mag respectively. 
This systematic effect is present, to a somewhat lesser extent, also in 
other studies of this type, although it was not noted by the authors. 
For example, in $\omega$ Cen  (Clement \& Rowe 2000) and in 
M5 (Kaluzny et al.\ 2000), if one considers only the RRab variables with 
$Dm<5$  and with no evidence of being ``longP/overluminous'', the width 
of the $M_V$(Fourier) distribution is a factor $\sim$1.7 smaller than that 
of the V distribution. In M3 this factor seems higher, up to $\sim$2.5.  
This ``compression'' effect cannot possibly be explained by reddening 
variations nor line-of-sight depth effects, particularly in M3, and casts 
serious doubts on the reliability of $<M_V>$ determinations with 
this method.  
A careful inspection of Fig. 2 in Jurcsik et al.\ (2003) can help find an 
explanation for this distortion effect. 
There, all the relevant Fourier parameters are plotted as a function 
of period for the RR Lyrae stars of M3. Since both $A_1$ and $A_3$ are well 
defined linear functions of period for the main body of the regular variables, 
they can be substituted in eq. (18) to retain only the dependence on period. 
We have estimated these relations numerically using the values of $A_1$ and 
$A_3$ listed in Table \ref{fparab}, and the above eq. (18) becomes 
then $M_V(RRab) \propto -0.166\log\,P$. The very weak dependence on 
period is what reduces the width of the $M_V$ distribution by losing the
connection with temperature and with the real luminosity distribution. 
A similar effect ($M_V(RRab) \propto -0.17P$) occurs if one uses the 
alternative formulation to derive $M_V$ as a function of $P$ (instead of 
$\log\,$P), $A_1$  and $\phi_{31}$ (Jurcsik 1998), as was done by Clement 
\& Rowe (2000) and Kaluzny et al.\ (2000), and seems to be at work with the 
RRc stars too, with a somewhat larger scatter. \\
ii) In Fig. \ref{cacciari.fig18} the values $M_V$(Fourier) derived from eq.s (17) 
and (18), transformed to $\log\,L$ with the help of the BC$_V$ values in 
Table \ref{phyabc}, are plotted vs the Fourier temperatures.  
One can see that the luminosities of both the RRc and RRab stars approximately 
agree with the luminosity level of the ZAHB, albeit with an unrealistically 
small scatter. 
However, these distributions are inconsistent with the requirements of the 
pulsation theory represented by the boundaries of the instability strip, 
mainly because of the distorted temperature distributions (cf. Sect. 6.2.2). 

\noindent $\bullet$ {\bf Blazhko variables} \\
We have applied eq. (18) also to the Blazhko stars separately at the various 
Blazhko phases, and we list the results in Table \ref{phyfbla}.  
The average of these values is $<M_V>$=0.53$\pm$0.05 mag at the large 
amplitude phase, 0.59$\pm$0.05 mag at the small amplitude phase, 
and 0.47$\pm$0.02 mag for the  5 evolved/overluminous stars at the 
large amplitude phase. We note that these values are comparable with those 
derived for the  RRab stars, so there seems to be no significant systematic 
difference in average luminosity between the regular and the Blazhko variables, 
as already noted in Sect. 3.1. 
 
We conclude that these relations based on the Fourier parameters may be of 
some use to estimate the average $M_V$ of a group of stars, when 
applicable and after proper calibration, but cannot be trusted to yield 
{\em accurate individual} $M_V$ estimates.

\subsubsection{The Mass} 

Based on a grid of hydrodynamic pulsation models for RRc variables at various 
masses, SC93 showed (their Fig. 2) that there is a clear relation between 
period and $\phi_{31}$ that depends essentially on mass, hence they derived 
a relation to estimate the mass for RRc stars: 

\noindent $\log\,M(RRc) = 0.52\log\,P - 0.11\phi_{31} + 0.39$ \hfill (19) 

\noindent The zero-point of this relation is such that intermediate metallicity 
clusters of Oosterhoff type I (e.g. M3, M5, NGC 6171) have RRc variables 
with average mass around 0.6 $M_{\odot}$, and metal-poor clusters of 
Oosterhoff type II (e.g. M68 and M15) have RRc variables with average mass 
around 0.8 $M_{\odot}$, in agreement with the values of mass for double-mode 
pulsators that were available and generally accepted at that time. 
However, the most recent estimates of mass for double-mode pulsators 
in M3 would support larger values by $\sim 0.10-0.15 M_{\odot}$, as well as  
a nearly flat dependence on metallicity (Clementini et al.\ 2004, 
cf. Sect. 4.3.1).  
   
We have listed in Table \ref{phyfc} the mass values derived from eq. (19), 
and we compare them in Fig. \ref{cacciari.fig18} with a few ZAHB models.  
In addition to the clumpy distribution due to  
the distorted Fourier temperatures, the mass distribution shows the opposite 
trend with respect to the theoretical distribution. 
 
On a different absolute scale,  Jurcsik (1998) provides a relation to derive 
the mass of RRab-type variables: 

\noindent $\log\,M(RRab)=20.884-1.754\log\,P+1.477\log\,L-6.272\log\,T_{eff}
+0.0367[Fe/H]$ \hfill (20) 

\noindent where the values of luminosity, metallicity and temperature 
have been taken from Table \ref{phyfab} (all derived from Fourier 
coefficients via 
eq. 18, 10 and 13 respectively, for self-consistency). 
We have listed in Table \ref{phyfab} the mass values derived from eq. (20). 
Also this mass distribution, like the RRc's, shows the opposite trend 
with respect to the  ZAHBs reported in Fig. \ref{cacciari.fig18}, in addition to the 
distortion due to the temperature distribution. 
We conclude that the Fourier Transform approach, in its present formulation, 
does not provide reliable values of mass for the RR Lyrae stars.

\subsubsection{The Gravity} 

Once the basic physical parameters mass, luminosity and temperature are 
known, the gravity can be calculated, for the sake of completeness, simply 
using the equation of the stellar structure:

\noindent $\log\,g=-10.607+\log\,M-\log\,L+4\log\,T_{eff}$ \hfill (21)

\noindent where $M$ and $L$ are in solar units. 
As a consequence of the mass distributions discussed above, we can see in 
Fig. \ref{cacciari.fig18} that the gravity distributions also fail to match the 
ZAHBs.
 
As a final comment, we stress again that the use of Fourier coefficients to 
estimate the physical parameters of the RR Lyrae stars, that might produce 
acceptable {\em average} results in some cases, e.g. with metallicity or 
luminosity after careful and proper calibration, is not presently able to 
provide reliable estimates of intrinsic colors hence temperatures and 
temperature-related parameters.

\section{Summary and Conclusions} 

We have performed a detailed study of the pulsational 
and evolutionary characteristics of 133 RR Lyrae variables in M3, 
selected among those with the best quality light curves from the 
CC01 data set. The availability of additional data sets (Car98 and Kal98) 
at different epochs has allowed us to study in good detail the 
characteristics of the Blazhko stars. Mean magnitudes and colors, along 
with periods, light curve amplitudes and rise times, have been used to 
discuss the pulsational properties of these stars. 
A critical discussion of the 
temperature determination process (i.e. temperature indicators and  
calibrations) has been presented, and the physical parameters and 
evolutionary characteristics of these stars have been estimated. 
The unusual richness of RR Lyrae stars in M3 and the excellent quality of 
the available data has allowed us to identify a good number of stars in 
a more evolved stage of evolution off the ZAHB and study their 
characteristics. Finally, we have performed a Fourier analysis of the V 
light curves and estimated the {\em pros} and {\em cons} of this technique 
when applied to the study of RR Lyrae properties. 

Our main conclusions are the following:   

\noindent $\bullet$ 
The basic characteristics of the CMD already discussed by CC01 are here 
reconfirmed, namely i) the blue and red edges of the instability strip are 
located at (B--V)=0.18 and 0.42, respectively; the RRc and RRab stars overlap 
in color in the interval $\sim$ 0.24 to 0.30. ii) The $<V>$ distribution 
is bimodal, with a main peak around $<V>$=15.64 and a secondary peak around
$<V>$=15.52. There is no significant evidence of {\em four} populations as 
claimed by Jurcsik et al.\ (2003). The intrinsic magnitude thickness of the 
HB within the instability strip is $\le$0.20 mag if we consider only 
the main (fainter) component, or $\sim$0.30 mag if we include also the 
brighter one. 
  
\noindent $\bullet$ 
At least one third of the RR Lyrae stars in M3 are affected by Blazhko 
modulation; 
in the studied sample, they all belong to the RRab group. More can be hidden 
in the sample we have not taken into account in the present analysis because 
of large scatter in the light curves. 
The presence of unidentified Blazhko stars causes a scatter in the 
relations among various observable parameters, that may be large enough to 
hide the presence of sub-groups with different characteristics. 
The properties of Blazhko stars at the Blazhko phase corresponding to the 
largest light curve amplitude are generally more similar to the 
characteristics of regular RRab stars than at smaller amplitude phases.  
The average $<V>$ magnitude does not vary significantly with Blazhko phase. 
The $<V>$ magnitude distribution of the Blazhko stars is the same as that 
of the regular RRab stars, including the bimodal shape. 
The average $<B-V>$ color distribution is also similar to the RRab's, but 
is truncated at a bluer color, i.e. there are no Blazhko stars redder than 
$<B-V> \sim$ 0.39.

\noindent $\bullet$
In the period-amplitude diagram both RRc and RRab stars are located on well 
defined sequences, that are more accurately represented by quadratic rather 
than linear relations (especially the sequence of the RRab stars), in agreement 
with theoretical models. There is clear evidence of nearly parallel sequences 
for both RRc and RRab stars, shifted towards longer periods and populated by 
systematically brighter stars than the respective main stellar groups. 
From our sample of 133 RR Lyraes we have identified 19 such stars (9 RRab, 
5 Blazhko and 5 RRc), that are all consistent with a more advanced stage 
of evolution off the ZAHB. Their distributions are similar to the mean
distributions of OoII RRc and RRab variables. The dependence of the 
P-A$_V$ relation on Oosterhoff type and/or evolutionary status rather than 
metallicity supports the conclusion that the Oosterhoff dichotomy is due 
to evolution. The numbers of RR Lyrae stars we have found in M3 near the 
ZAHB and evolved off the ZAHB are consistent with evolutionary lifetimes 
according to well established theoretical considerations. 
One of the three shortest period and lowest amplitude RRc stars is likely 
to be a second overtone pulsator. 

\noindent $\bullet$
After a critical discussion of what is the most reliable mean color as an 
indicator of the equivalent static color for an RR Lyrae star, we have 
decided to use the formulation $(B-V)_S=<B>_{int}-<V>_{int}$ plus 
amplitude related corrections based on theoretical models (Bono et al.\ 1995) 
that are quite consistent also with empirical estimates (Sandage 1990). 
From these colors and using a few independent methods we have estimated 
a mean reddening  E(B--V)=0.01$\pm$0.01 for M3. 
A comparative evaluation of various temperature scales has led us to
identify two temperature scales that meet both theoretical (pulsational and 
evolutionary) requirements and observational evidence on mass and luminosity 
for the RR Lyrae stars using (B--V) colors (in absence of V--K colors). 
These scales are from M98 (theoretical calibration) and SF. They differ by 
$\sim$150 K (M98 being cooler), and yield on average pair values of 
mass(M$_{\odot}$)/M$_V$(mag) about 0.74/0.59 and 0.69/0.54, respectively. 
The temperature scale by CSJ is very similar to SF's and is independent of
color, but is defined only for RRab stars. 
Considering that the M98 calibrations based on the (V--K) colors are supposed 
to be more reliable and are both $\sim$100 K hotter than the corresponding 
calibrations based on (B--V) colors, we have adopted the hotter temperature 
scale by SF for our analysis (corresponding to a distance modulus of 15.07 
for M3).  
However, our considerations would hold also with the cooler (B--V)-based M98 
scale and a distance modulus of 15.02, within the errors. 
By using the SF temperature scale and the (B--V)$_S$ colors we have  
derived the stellar physical parameters (temperature, luminosity, mass and 
gravity) for our stars, and compared them with the most recent stellar 
evolution and pulsation models. The agreement is good, and confirms that the 
adopted calibration is reliable and accurate, and yields fully consistent 
results with the theoretical framework within the respective errors. The use 
of the CSJ temperature scale yields equally good or better (less dispersed) 
results, for the RRab variables only. 
 
\noindent $\bullet$
We have applied the Fourier Transform technique to our variables.  
The main aim was to exploit our excellent data set and investigate  
the reliability of this type of analysis. 
First, we have derived the $Dm$ parameter, defined by Jurcsik \& Kov\`acs 
(1996) as a quality indicator of the regularity of the light curve shape. 
Only for $Dm<3$ the physical parameters derived from Fourier coefficients are 
considered ``reliable'', according to Kov\`acs \& collaborators' prescriptions.  
We have adopted $Dm<5$ to increase the statistics with no significant 
loss of accuracy. We have found that $Dm$ is effectively unable to 
distinguish between Blazhko and non-Blazhko stars unless set to an 
unpractically low value ($Dm \le 2$). Even among Blazhko stars, one can find 
the recommended value $Dm<3$ as frequently at small-amplitude as at 
large-amplitude Blazhko phases. 
About the Fourier analysis results (for stars with $Dm<5$), we have found the 
following: \\ 
i) [Fe/H] estimates seem on average acceptable, but are $\sim$0.1 dex more 
metal-rich than the high resolution spectroscopic abundances of red giant 
stars derived by KI. 
A recalibration of the [F/H]-period-$\phi_{31}$  relation using 287 RRab
variables in 18 globular clusters performed by Sollima et al.\ (2004) using a 
non-parametric fitting method and KI metallicity scale yields similar average 
metallicity values to those derived from eq. (10), within the errors. 
This indicates that M3 lies $\sim$0.1 dex off (on the metal-poor side) the 
mean relation defined by the calibrating globular clusters.   
However, the use of a different fitting method and/or metallicity 
scale, such as Sollima et al.'s, produces a different {\em shape} of the 
metallicity distribution, that might become relevant in the case of a composite 
population with a non-negligible metallicity dispersion. 
Evolution off the ZAHB does not affect [Fe/H] determinations. The 
inclusion of Blazhko stars in a sample of regular stars does increase the 
scatter in the [Fe/H] determinations, as Blazhko stars at low amplitude phase 
appear as more metal-rich. If this effect is taken into account, there is no 
evidence of metallicity spread among the RR Lyrae stars in M3. \\
ii) Intrinsic colors and temperatures estimated from eq.s (11)-(15) 
show serious discrepancies with observed color distributions and theoretical 
(pulsational and evolutionary) requirements, and cannot be taken as reliable 
results. \\
iii) Absolute magnitudes are affected by a  ``compression'' effect that reduces 
their scatter by a factor $\sim$2 compared to the observed $<V>$ distribution. 
This makes them unreliable as {\em accurate individual} values, but they may 
provide useful averages for groups of stars, if applicable and after proper 
calibration. The r.m.s. errors are however significantly underestimated. \\
iv) The values of mass show a distribution with temperature that has the 
opposite trend with respect to the ZAHB. Consequently, also the values of 
gravity are affected by serious uncertainties. Neither estimates can be taken 
as reliable results.   

In general, it appears that the physical parameters of RR Lyrae stars derived 
from the Fourier decomposition of the V light curves should be taken with 
considerable caution. In particular, intrinsic colors hence temperatures and 
temperature-related parameters (e.g. mass and gravity) are seriously 
inaccurate. Possible exceptions are [Fe/H], that seem to be acceptable
within r.m.s. errors of $\pm$0.15 dex (provided Blazhko stars are not included 
in the analysis and only average values are considered), and absolute 
magnitudes if taken as the mean value for a group and not as individual values.

\acknowledgements 

We wish to thank T.D. Kinman for providing the calibration constant for our 
eq. (18); A.V. Sweigart for providing his HB models in electronic form and for 
interesting discussions and comments; D. VandenBerg for making available his 
ZAHB models; A. Sandage for sending his analysis on 
Fourier parameters in advance of publication, and inducing us to check the 
effect of $\phi_{31}$  on metallicity determinations; and G. Clementini for 
several interesting discussions and comments.
We acknowledge the use of J. Kaluzny original data that are available on the 
web. 
TMC wishes to thank the UNC Charlotte for a Reassignment of Duties Grant.
BWC thanks the US National Science Foundation for grants AST-9619881,
AST-9988156, and AST-0305431 to the University of North Carolina. 
CC aknowledges the support by the MIUR (Ministero dell'Istruzione, 
dell'Universit\`a e della Ricerca).

\clearpage



\clearpage


\clearpage

\clearpage

\clearpage

\clearpage

\clearpage

\clearpage

\clearpage

\clearpage

\clearpage

\clearpage

\clearpage

\clearpage

\clearpage

\clearpage

\clearpage

\clearpage

\clearpage

\clearpage

\end{document}